\documentclass[lettersize,journal]{IEEEtran}
\usepackage{color,array,amsthm}
\usepackage{graphicx}
\usepackage{amsmath,amsfonts,hyperref,soul, color, xcolor}
\usepackage{algpseudocode}
\usepackage{CJKutf8} 
\usepackage{algpseudocode}
\usepackage{algorithm}
\usepackage{array}
\usepackage{textcomp}
\usepackage{multirow}
\usepackage{stfloats}
\usepackage{amssymb}
\usepackage{lipsum}
\usepackage{cuted}
\usepackage{url}
\usepackage{booktabs} 
\usepackage{verbatim}
\usepackage{cite}
\usepackage{caption}
\usepackage{subcaption}
\usepackage{amsthm} 
\usepackage{etoolbox}
\patchcmd{\thebibliography}{\section*{\refname}}{}{}{}

\theoremstyle{definition}

\newtheorem{theorem}{Theorem}

\theoremstyle{remark}

\usepackage{siunitx} 
\DeclareMathOperator{\atan}{atan}
\DeclareMathOperator{\asin}{asin}
\DeclareMathOperator{\atanTwo}{atan2}

\begin{document}

\title{Low-Cost Infrastructure-Free 3D Relative Localization with Sub-Meter Accuracy in Near Field} 

\author{Qiangsheng~Gao,~\IEEEmembership{Member,~IEEE,}
Ka~Ho~Cheng,~
Li~Qiu,~\IEEEmembership{Fellow,~IEEE,}~\\
and~Zijun~Gong,~\IEEEmembership{Member,~IEEE}
\thanks{This work was supported by HKUST-Bright Dream Robotics Joint Research Institute (HBJRI) Fund Scheme under grant HBJRI-FTP-004 and Shenzhen Science and Technology Innovation Committee under Shenzhen-Hong Kong-Macau Science and Technology Program (Category C) SGDX20201103094600006. {\itshape (Corresponding author: Z. Gong)}}
\thanks{The authors were with the Department of Electronic and Computer Engineering, The Hong Kong University of Science and Technology, Hong Kong (email: qgaoac@connect.ust.hk; khchengah@connect.ust.hk; eeqiu@ust.hk; gongzijun@ust.hk). Z. Gong is also with the IOT Thrust, HKUST (Guangzhou), Guangzhou, Guangdong 511453, China. }}



\maketitle

\maketitle

\begin{abstract}
Relative localization in the near-field scenario is critically important for unmanned vehicle (UxV) applications. Although related works addressing 2D relative localization problem have been widely studied for unmanned ground vehicles (UGVs), the problem in 3D scenarios for unmanned aerial vehicles (UAVs) involves more uncertainties and remains to be investigated. Inspired by the phenomenon that animals can achieve swarm behaviors solely based on individual perception of relative information, this study proposes an infrastructure-free 3D relative localization framework that relies exclusively on onboard ultra-wideband (UWB) sensors. 
Leveraging 2D relative positioning research, we conducted feasibility analysis, system modeling, simulations, performance evaluation, and field tests using UWB sensors. The key contributions of this work include:
derivation of the Cram\'er-Rao lower bound (CRLB) and geometric dilution of precision (GDOP) for near-field scenarios;
development of two localization algorithms -- one based on Euclidean distance matrix (EDM) and another employing maximum likelihood estimation (MLE);
comprehensive performance comparison and computational complexity analysis against state-of-the-art methods;
simulation studies and field experiments;
a novel sensor deployment strategy inspired by animal behavior, enabling single-sensor implementation within the proposed framework for UxV applications.
The theoretical, simulation, and experimental results demonstrate strong generalizability to other 3D near-field localization tasks, with significant potential for a cost-effective cross-platform UxV collaborative system.
\end{abstract}

\begin{IEEEkeywords}Relative localization, ultra-wideband (UWB), Cram\'er-Rao lower bound (CRLB), range measurement, Euclidean distance matrix (EDM).
\end{IEEEkeywords}

\section{INTRODUCTION}
\label{sect:Introduction}
Precise localization is essential in diverse domains, including multi-agent robotic systems, the Internet of Things, intelligent vehicular networks, and logistics \cite{buehrer_collaborative_2018,evrendilek_complexity_2011,xu2025SurveyUltraWidea}. 
The low-altitude economy and unmanned vehicle (UxV) logistics are thriving with the swift progress and growing sophistication of the robotics industry and sensor technology. Due to the payload capacity and endurance limit of unmanned aerial vehicles (UAVs), researchers have explored novel delivery concepts involving heterogeneous vehicle cooperation approaches like ``trucks/unmanned ground vehicles (UGVs) + small UAVs'' \cite{madani2022HybridTruckDroneDelivery,attenni2023DroneBasedDeliverySystems,moadab2022DroneRoutingProblem}, ``large branch UAVs + small end UAVs'' \cite{WEN2022HeterogeneousMultidrone}, and ``UAVs + smart stations + couriers'' \cite{chen2025CoordinatedDroneCourierLogistics}. 
Logistic companies have already deployed UGVs for express delivery in large residential areas. Leading companies such as SF Express, Amazon, and Keeta have further integrated UAVs into their last-mile delivery systems \cite{WEN2022HeterogeneousMultidrone,chen2025CoordinatedDroneCourierLogistics}. This technology demonstrates broad applicability, including emergency medical supply transportation in mountainous rural regions and short-distance high-frequency time-sensitive deliveries of perishable goods, such as Yangcheng Lake hairy crabs.
However, current UAV transportation is facing challenges that need to be addressed, such as insufficient localization precision, inadequate adaptability to complex environments, low efficiency in cross-platform collaborations, high initial investment costs, limited payload capacity and battery endurance \cite{moadab2022DroneRoutingProblem}, complex airspace approval processes, security and privacy risks \cite{hahn2021SecurityPrivacyIssuesa,bendiab2023AutonomousVehiclesSecurity}, etc. This paper investigates the challenges of \textit{near-field 3D localization} in the global navigation satellite systems (GNSS)-denied environments, with a focus on heterogeneous UxV platforms.

Primarily speaking, localization can be classified into infrastructure-based localization and infrastructure-free localization depending on whether pre-set infrastructures are required \cite{gao_TIM_2025_InfrastructureFreeRelativeLocalization}. Localization can also be categorized into absolute localization which determines target positions in the global coordinate system, and relative localization which calculates target positions with respect to the onboard coordinate system.
Infrastructure-based absolute localization is exemplified by GNSS satellite localization, which is widely used due to its broad applicability and high accuracy. GNSS localization typically achieves accuracy in the $\SI{5}{\m}-\SI{10}{\m}$ order, but various factors including UAV mobility, insufficient satellite visibility, multipath effects, ionospheric delay, and electromagnetic interference can degrade accuracy to $\SI{10}{\m}-\SI{40}{\m}$ order in adverse environments such as mountains, canyons, and urban areas \cite{zhang2022EfficientUAVLocalization,lu2024FastOmnidirectionalRelative}.
Things get even worse for GNSS under circumstances like caves and indoor areas where satellite signals are shielded and GNSS cannot even be used. 
Moreover, GNSS provides better horizontal accuracy but suffers from poorer vertical accuracy, making it less suitable for UxV applications, especially those involving UAVs.
While horizontal localization accuracy is widely used for daily life and ground vehicles, the vertical accuracy plays an equally or even more significant role for UAV applications. Investigations have been conducted to improve the localization accuracy and the state-of-the-art method is by employing real-time kinematic (RTK) technology, which is effective and achieves $\SI{}{\cm}$ order accuracy. However, the price for RTK is relatively high, and it is hard to guarantee that all the UGVs are equipped with RTK. Vision-based localization has quite high accuracy in \SI{}{\cm} and even \SI{}{\mm} order, but its localization range is limited and it is severely influenced by external weather and light conditions \cite{zeng2023UAVLocalizationSystem,lazzari2017NumericalInvestigationUWB}.


On the other hand, animals can achieve complex collective behaviors, such as birds flying in V-shaped formations \cite{mirzaeinia2020AnalyticalStudyLeader}, fish swimming in vortexes \cite{ito2022EmergenceGiantRotating}, and zebras and wildbeest migrating in herds \cite{gueron1996DynamicsHerdsIndividualsa}, to improve their probability of survival in adverse environment. All these collective animal behaviors are done with each individual relying merely on the relative information from its own sensory organs and controlling the behavior of itself. Inspired by these natural animal behaviors, related research indicates that infrastructure-free relative localization simplifies multi-agent UxV collaboration in contrast to absolute localization and infrastructure-based relative localization \cite{de_silva_ultrasonic_2015,brunacci_development_2023,vasarhelyi_optimized_2018}. Therefore, an accurate low-cost infrastructure-free relative localization framework is required for UxV applications in the range order from $\SI{}{\dm}$ to $\SI{100}{\m}$. Ultra-wideband (UWB) range-measurement sensors are suitable for low-cost onboard applications due to their cost-effectiveness, light weight, low power consumption, and centimeter-level ranging accuracy.
Range-based localization has been studied by researchers for many years, but most of the work is in the 2D scenarios \cite{cao_relative_2021,gao_TIM_2025_InfrastructureFreeRelativeLocalization,fishberg_multi-agent_2022}.
Considering the higher degree-of-freedom and the inherently non-uniform spatial distribution of measurement errors in 3D scenarios, the UWB-based 3D localization faces many challenges.

In our work, the 3D infrastructure-free relative localization for a multi-UxV system is studied. System modelling, sensor configuration, and algorithm design are studied. Then we conducted performance analysis for different algorithms in simulations and field tests. The UxV application is then studied to validate the feasibility of our proposed framework for industry scenarios.

The contributions of this work are listed below. 
\begin{enumerate}
        \item The geometric dilution of precision (GDOP) and Cram\'er-Rao lower bound (CRLB) for 3D localization are derived and used as performance metrics. Then the minimum number of sensors is studied and the optimal sensor configuration is derived and adopted. 
        \item Two algorithms, the Euclidean distance matrix (EDM)-based trilateration (EDMT) algorithm and the probability-model-based maximum likelihood estimation (MLE) algorithm, are proposed for 3D scenarios to localize a target sensor and a target agent. The root mean square error (RMSE) of the predicted position closely aligns with the CRLB when applying the proposed algorithms. 
        \item The computational complexity benchmarks and accuracy benchmarks in this study encompass two state-of-the-art methods, traditional trilateration (TT) and Frobenius-norm-based estimation (Fro-CVX), respectively. The RMSEs of the proposed methods are contrasted with those of TT and Fro-CVX, alongside the CRLB. The theoretical computational complexity for all methods are analyzed and compared, and the real computational time is documented for each method. 
        \item Experimental and simulation results are showcased and scrutinized to validate the theoretical outcomes. The field tests further confirm the feasibility of the suggested framework in 3D environments. Inspired by the bee's position information transmission method, the waggle dance \cite{dong2023SocialSignalLearning}, the application for 3D UxV localization is then presented with only one UWB sensor configured on each UxV. 
\end{enumerate} 
In the following, Section II introduces the preliminary and methodology for 3D relative localization, as well as the basics of the proposed methods and the state-of-the-art methods. Section III derives the algorithms in detail for 3D scenarios and the simulations are studied in Section IV. Then we demonstrate the field test result and the application adaptability in Section V and get the conclusion in Section VI. 

\textit{Notation: } In this article, we adopt the following notation conventions: italic lowercase, bold italic lowercase, and uppercase letters like $x$, $\boldsymbol{p}$ and $A$, are used to represent scalars, vectors, and matrices, respectively. The transpose of a matrix $A$ is denoted as $A^{\top}$, the trace of a square matrix $A$ is denoted as $\mathrm{trace}(A)$, and the Frobenius norm of a real matrix $A$ is defined as $\|A\|_F\triangleq\sqrt{\mathrm{trace}(A^{\top}A)}$. Let $I_n$ be the $n$-dimensional identity matrix and $\mathbf{1}_{n}$ (respectively $\mathbf{0}_{n}$) the $n$-dimensional column vector with all entries of $1$ (respectively $0$). A Gaussian random variable $\epsilon$ with a mean $\mu$ and a variance $\sigma^2$ is denoted as $\epsilon\sim\mathcal{N}(\mu,\sigma^2)$. 

\section{METHODOLOGY}
\label{sect:Methodology}
In this section, we discuss the infrastructure-free 3D relative localization problem using only the minimum required range-measurement information \cite{shang_localization_2003,cao_sensor_2006,cao_formation_2011}. The GDOP and CRLB are derived and used as performance metrics, similar to the 2D cases in \cite{gao_TIM_2025_InfrastructureFreeRelativeLocalization}. Then we clarify the minimum required number of sensors and compare the average and maximum GDOP in different settings to determine the optimal sensor configuration for a 3D agent. Next, two commonly used algorithms, TT and Fro-CVX, are presented and will be used as benchmarks for computational complexity and measurement accuracy, respectively. Finally, the basics of the proposed algorithms, the EDMT and MLE, are introduced. 

\subsection{Localizability}
\label{subsect:localizability}
Infrastructure-free relative localization has its advantages for multi-agent cooperation and is investigated by researchers for years \cite{gao_TIM_2025_InfrastructureFreeRelativeLocalization,fishberg_multi-agent_2022}. Different from the UGV applications where we can usually simplify the mathematical models to 2D scenarios, applications with UAV involved is mainly in 3D scenarios. Therefore, the localizability in 3D scenarios should be discussed.

\begin{figure}[!htbp]
 \centering
 \includegraphics[height=0.3\textwidth]{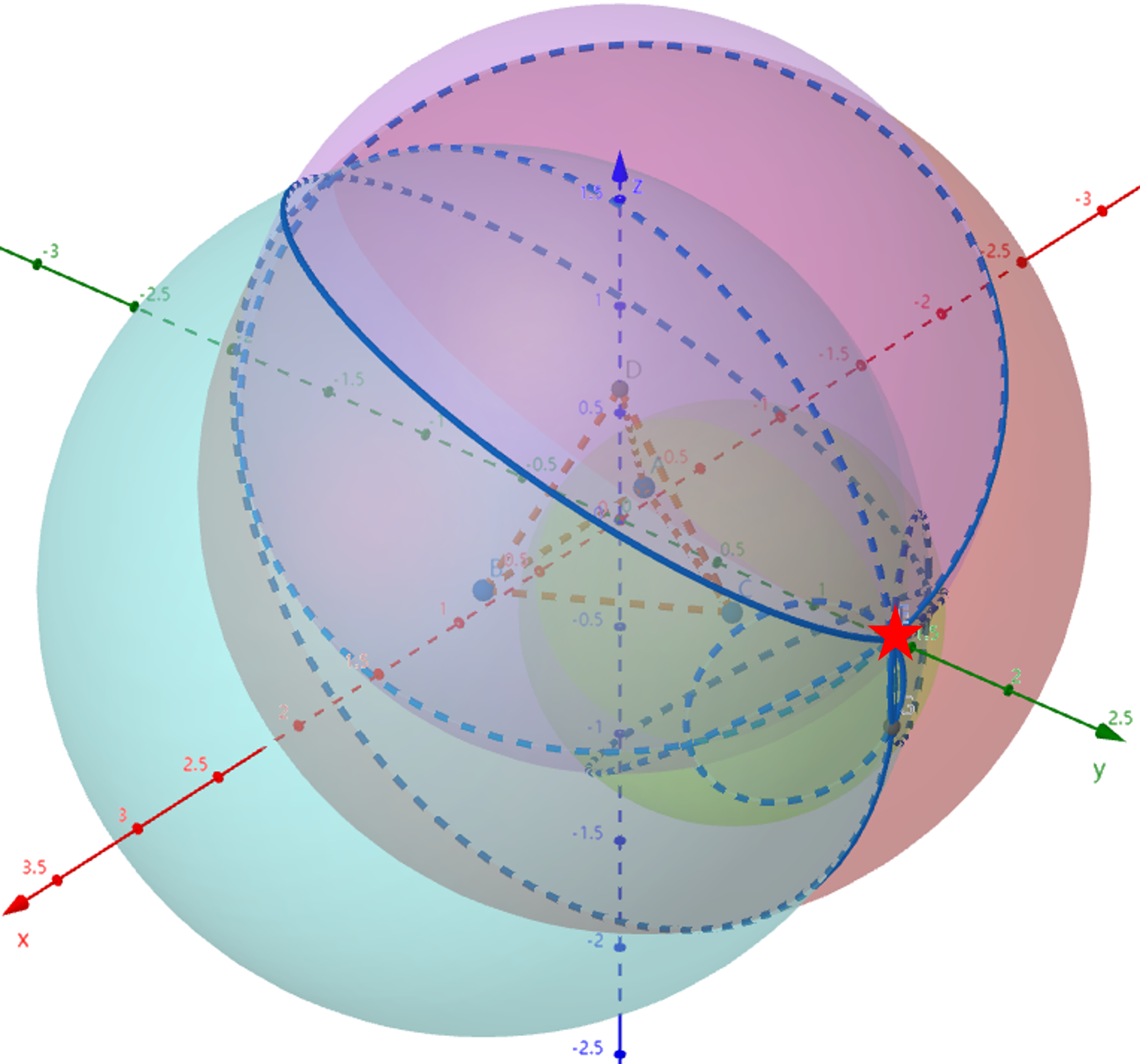}
 \caption{Unambiguous relative localization with four ranging sensors. }
 \label{fig:UnambiguousPointLocalization_4xRanges_3D}
\end{figure}

Here we discuss the 3D localizability for a UxV with onboard range-measurement sensors. Each measured distance constrains the target to a sphere centered at the onboard sensor. As illustrated in Fig.~\ref{fig:UnambiguousPointLocalization_4xRanges_3D}, we need to use at least four sensors for 3D relative localization, where any three of which are not colinear and any four of which are not coplanar, to uniquely determine the target position.



\subsection{Preliminary}
For the convenience of discussion, we will first introduce the roll, pitch and yaw angles and their 3D rotation matrix. Then the notations of the state of the target, the real distance vectors, and measured distance vectors will be presented to localize a target sensor and a target agent.
\subsubsection{3D Rotation Matrix}
\begin{figure}[!htbp]
  \centering
  \includegraphics[width=0.3\textwidth]{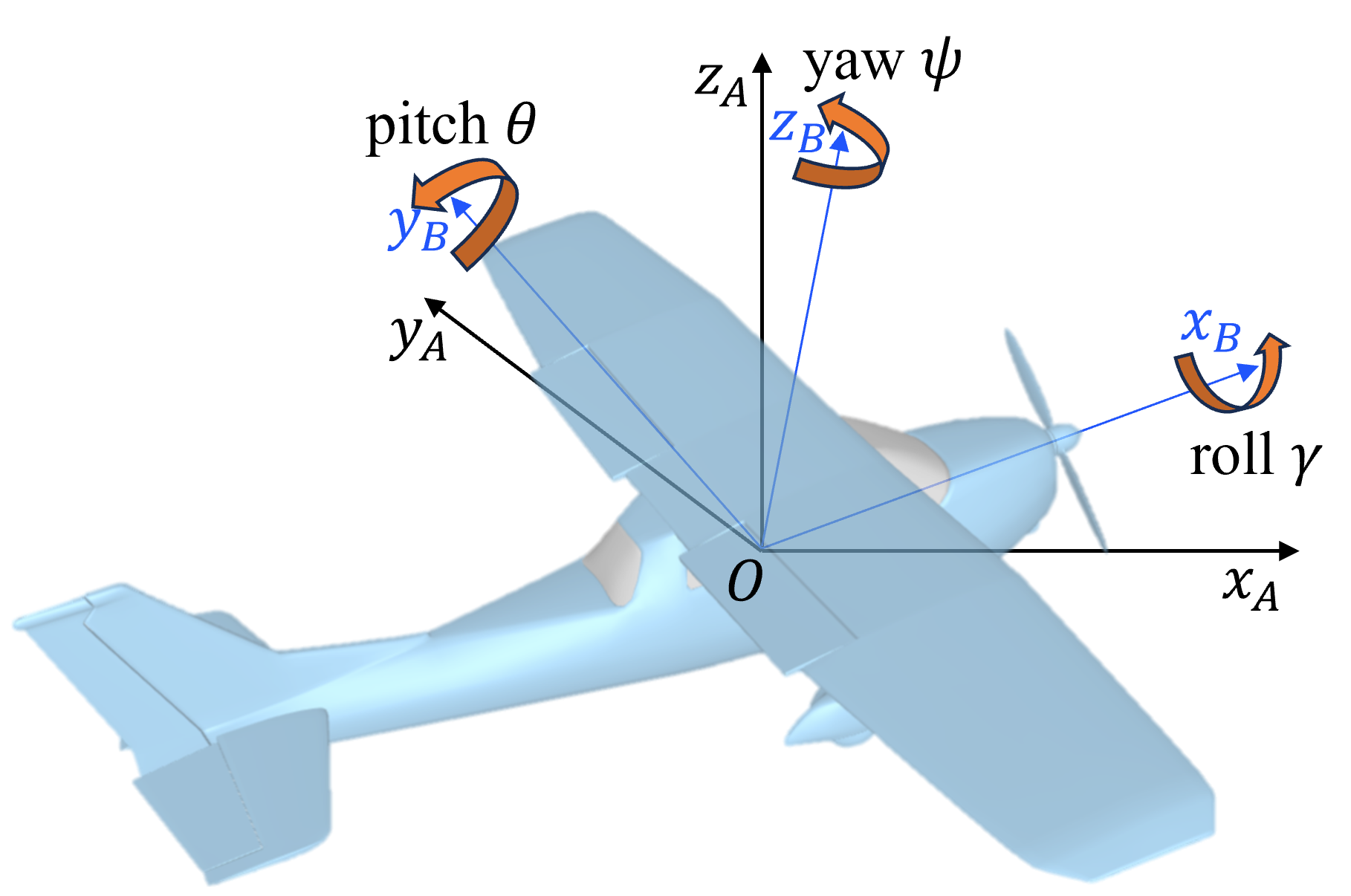}
  \caption{Roll, pitch, and yaw rotation angles $\gamma,\theta,\psi$ about axis $x$, $y$, and $z$.}
  \label{fig:YawPitchRoll_Demo_3D}
\end{figure}

In 3D scenarios, there are three rotation matrices corresponding to rotation about three different axes. The $x$, $y$ and $z$ axes of the body frame corresponding to roll, pitch, and yaw rotation angles $\gamma$, $\theta$, and $\psi$ are shown in Fig.~\ref{fig:YawPitchRoll_Demo_3D}. 

The corresponding basic rotation matrices, $R_1(\gamma),R_2(\theta),R_3(\psi)$, are defined by 
\begin{equation}
 R_1(\gamma)\triangleq\begin{bmatrix}
        1 & 0 & 0 \\ 
        0 & \cos\gamma & -\sin\gamma \\ 
        0 & \sin\gamma & \cos\gamma
\end{bmatrix}, 
\label{eq:3D_RotationMatrix_R1}
\end{equation}
\begin{equation}
R_2(\theta)\triangleq\begin{bmatrix}
        \cos\theta & 0 & \sin\theta \\
        0 & 1 & 0 \\
        -\sin\theta & 0 & \cos\theta
\end{bmatrix}, 
\label{eq:3D_RotationMatrix_R2}
\end{equation}
\begin{equation}
R_3(\psi)\triangleq\begin{bmatrix}
        \cos\psi & -\sin\psi & 0 \\
        \sin\psi & \cos\psi & 0 \\ 
        0 & 0 & 1
\end{bmatrix}.
\label{eq:3D_RotationMatrix_R3}
\end{equation}
It is important to note that the order of rotations around different axes can be approximately ignored only when the rotation angles are sufficiently small. When the rotation angles are not all small, the coordinate transformation matrix is dependent on the order of rotations around different axes. For identical rotation matrices but different rotation orders, different coordinates are produced, which demonstrates the non-commutativity of finite rotations.

In subsequent discussions, we will rotate the coordinate system in the order of yaw, pitch, and roll, i.e., the rotation matrix is 
\begin{equation}
\mathbf{R}(\gamma,\theta,\psi) = \mathbf{C}_{A}^{B}(\gamma,\theta,\psi) = R_{1}(\gamma)R_{2}(\theta)R_{3}(\psi) 
\label{eq:YawPitchRoll_RotationMatrix}
\end{equation}
Then we have 
$$\begin{bmatrix} x_B \\ y_B \\z_B\end{bmatrix}=\mathbf{C}_{A}^{B} \begin{bmatrix} x_A \\ y_A \\ z_A \end{bmatrix}.$$ 
Once we get the rotation matrix $$\mathbf{R}(\gamma,\theta,\psi) = \mathbf{C}_{A}^{B} = \begin{bmatrix}
  R_{11} & R_{12} & R_{13} \\ R_{21} & R_{22} & R_{23} \\ R_{31} & R_{32} & R_{33} 
\end{bmatrix},$$ 
we can also get the corresponding yaw, pitch, and roll angles from $\mathbf{R}(\gamma,\theta,\psi)$ by 
\begin{equation}
\begin{aligned}
    \psi&=\atanTwo\left({-R_{12}},{R_{11}}\right),\\
    \theta&=\asin R_{13}, \\
    \gamma&=\atanTwo\left({-R_{23}},{R_{33}}\right),
\end{aligned}
\label{eq:Get_YawPitchRollfromR}
\end{equation}
where the 2-argument arctangent function $\atan2(y, x)$ computes the angle between the positive $x$-axis and the point $(x, y)$ in the Cartesian plane and the range is $(-\pi,\pi]$. 
\subsubsection{Target States and Distance Vectors}
In the following, the positions of the onboard sensors with respect to its own coordinate system are denoted by $\boldsymbol{p}_i=\begin{bmatrix} x_i & y_i & z_i \end{bmatrix}^{\top}$, where $i=1,2,\ldots,k$, and $k$ is the number of sensors onboard and equals four in our subsequent setting. The positions of the sensors on agent-$A$ are denoted by $\boldsymbol{p}_{A_i}$, where $i=1,2,\ldots,k$. Obviously, $\boldsymbol{p}_{A_i}=\boldsymbol{p}_i$ for any $i=1,2,\ldots,k$. Then we investigate the relative localization problem to localize a sensor and an agent from the coordinate system of agent-$A$. 

To localize a target sensor, the real state of target sensor $T$ is denoted as $\boldsymbol{p}_T=\begin{bmatrix} x_T & y_T & z_T\end{bmatrix}^{\top}$, where $x_T,y_T,z_T$ are the $x$, $y$, and $z$ coordinates with respect to the coordinate system of agent-$A$, and the estimate of $\boldsymbol{p}_T$ is denoted as $\boldsymbol{\hat{p}}_T=\begin{bmatrix} \hat{x}_T &\hat{y}_T & \hat{z}_T\end{bmatrix}^{\top}$. 
The real distance vector from onboard sensor $A_i$ to the target is a function of $\boldsymbol{p}_T$, i.e., 
\begin{equation}
\boldsymbol{d}_{p}(\boldsymbol{p}_T)=\begin{bmatrix}
{d}_{A_1T}(\boldsymbol{p}_T)&{d}_{A_2T}(\boldsymbol{p}_T)&\cdots&{d}_{A_kT}(\boldsymbol{p}_T)
\end{bmatrix}^{\top},
\label{eq:vector_d_I_3D}
\end{equation}
where 
\begin{equation}
 d_{A_iT}(\boldsymbol{p}_T)=\sqrt{(x_{A_i}-x_T)^2+(y_{A_i}-y_T)^2+(z_{A_i}-z_T)^2}.
 \label{eq:d_AiT_3D}
 \end{equation} 
The corresponding measured distance vector is denoted as 
\begin{equation}
 \hat{\boldsymbol{d}_{p}}=\begin{bmatrix}
\hat{d}_{A_1T}&\hat{d}_{A_2T}&\cdots&\hat{d}_{A_kT}
\end{bmatrix}^{\top},
 \label{eq:vector_hat_d_I_3D}
 \end{equation} where $\hat{d}_{A_iT}=d_{A_iT}+\epsilon_i,i=1,2,\ldots,k$, and $\epsilon_i$ is the measurement error, assumed to be independent and identically distributed (i.i.d.) Gaussian. 

To localize an agent, without loss of generality, we only consider the situation when $k=4$. The state of the target agent-$B$ is denoted as 
\begin{equation}
\boldsymbol{\beta}_{{B}}=\begin{bmatrix} x_{B}&y_{B}&z_{B}&\gamma_{B}&\theta_{B}&\psi_{B} \end{bmatrix}^{\top},
\label{eq:beta_B_3D}
\end{equation} where $x_B,y_B,z_B$ are the $x$, $y$, and $z$ coordinates, and $\gamma_{B},\theta_{B},\psi_{B}$ are the roll, pitch, and yaw angles with respect to agent-$A$, respectively. Also, denote the sensor position on agent-$B$ as $\boldsymbol{p}_{B_j}=\begin{bmatrix} x_{B_j} & y_{B_j} & z_{B_j}\end{bmatrix}^{\top}$, where $j=1,\ldots,4$. 
The estimate of $\boldsymbol{\beta}_{B}$ is denoted as $$\boldsymbol{\hat{\beta}}_{B}=\begin{bmatrix} \hat{x}_{B}&\hat{y}_{B}&\hat{z}_{B}&\hat{\gamma}_{B}&\hat{\theta}_{B}&\hat{\psi}_{B} \end{bmatrix}^{\top},$$ and the estimate of $\boldsymbol{p}_{B_j}$ is denoted as 
$$\boldsymbol{\hat{p}}_{B_j}=\begin{bmatrix} \hat{x}_{B_j} & \hat{y}_{B_j} &\hat{z}_{B_j}\end{bmatrix}^{\top},$$ $j=1,2,3,4$. For notation convenience, we define the real and estimated target agent positions as $$\boldsymbol{p}_B=\begin{bmatrix} x_B \\ y_B \\ z_B\end{bmatrix}\triangleq\frac{1}{4}\sum\limits_{j=1}^4\boldsymbol{p}_{B_j},\boldsymbol{\hat{p}}_B=\begin{bmatrix} \hat{x}_B \\ \hat{y}_B \\\hat{z}_B \end{bmatrix}\triangleq\frac{1}{4}\sum\limits_{j=1}^4\boldsymbol{\hat{p}}_{B_j}.$$ 
The real distance vector $\boldsymbol{d}_{B}$ from agent-$A$ to the target agent-$B$ is a function of $\boldsymbol{\beta}_B$, i.e., 
$$\boldsymbol{d}_{B}(\boldsymbol{\beta}_B)=\begin{bmatrix}
\boldsymbol{d}_{p}^{\top}(\boldsymbol{p}_{B_1}) & 
\boldsymbol{d}_{p}^{\top}(\boldsymbol{p}_{B_2}) &
\boldsymbol{d}_{p}^{\top}(\boldsymbol{p}_{B_3}) &
\boldsymbol{d}_{p}^{\top}(\boldsymbol{p}_{B_4}) 
\end{bmatrix}^{\top},$$ 
where $\boldsymbol{d}_{p}(\boldsymbol{p}_{B_j}),j=1,\ldots,4$ is defined previously in Eq. (\ref{eq:vector_d_I_3D}), $\boldsymbol{p}_{B_j}$ is a function of $\boldsymbol{\beta}_{B}$ with 
\begin{equation}
\boldsymbol{p}_{B_j}(\boldsymbol{\beta}_{{B}}) = \begin{bmatrix} x_{B_j}(\boldsymbol{\beta}_{{B}})\\y_{B_j}(\boldsymbol{\beta}_{{B}})\\ z_{B_j}(\boldsymbol{\beta}_{{B}})\end{bmatrix}=\begin{bmatrix} x_{B}\\y_{B}\\z_{B} \end{bmatrix}
 + \mathbf{C}_{B}^{A} \boldsymbol{p}_{j}, 
\label{eq:q_Bj_3D}
\end{equation}
where $j=1,2,3,4$, and 
\begin{equation} \label{eq:C_B^A}
\begin{split}
\mathbf{C}_{B}^{A}(\gamma,\theta,\psi)&=(\mathbf{C}_{A}^{B})^{-1}(\gamma,\theta,\psi)=(\mathbf{C}_{A}^{B})^{\top}(\gamma,\theta,\psi)\\
&=(R_{1}(\gamma_{B})R_{2}(\theta_{B})R_{3}(\psi_{B}))^{\top}\\
&=R_3(\psi_{B})^{\top}R_2(\theta_{B})^{\top}R_1(\gamma_{B})^{\top}\\
&=R_3(-\psi_{B})R_2(-\theta_{B})R_1(-\gamma_{B}).
\end{split}
\end{equation} 
Here $R_1(-\gamma_{B}),R_2(-\theta_{B}),R_3(-\psi_{B})$ are the basic 3D rotation matrices corresponding to rotation about $x$, $y$, and $z$ axis defined in Eq. (\ref{eq:3D_RotationMatrix_R1}), Eq. (\ref{eq:3D_RotationMatrix_R2}), and Eq. (\ref{eq:3D_RotationMatrix_R3}). 
The corresponding measured distance vector is denoted as 
$$\boldsymbol{\hat{d}}_{B}=\begin{bmatrix}
\boldsymbol{\hat{d}}_{p}^{\top}(\boldsymbol{p}_{B_1}) & 
\boldsymbol{\hat{d}}_{p}^{\top}(\boldsymbol{p}_{B_2}) &
\boldsymbol{\hat{d}}_{p}^{\top}(\boldsymbol{p}_{B_3}) &
\boldsymbol{\hat{d}}_{p}^{\top}(\boldsymbol{p}_{B_4}) 
\end{bmatrix}^{\top},$$ where $\boldsymbol{\hat{d}}_{p}(\boldsymbol{p}_{B_j}),j=1,\ldots,4$ is defined in Eq. (\ref{eq:vector_hat_d_I_3D}), and $\hat{d}_{A_iB_j}=d_{A_iB_j}+\epsilon_{i,j},i,j=1,\ldots,4$ is the measurement error $\epsilon_{i,j}$, assumed to be i.i.d. Gaussian. 

\subsection{GDOP and CRLB}
\label{subsect:GDOP_and_CRLB_3D}
Compared with 2D relative localization, deteriorated distribution of measurement error, more complex multipath effect, and higher data processing complexity occur in 3D scenarios. Therefore, the precise localization of a target in 3D space faces more challenges. In this subsection, two performance metrics, GDOP and CRLB, are derived to evaluate the accuracy of different sensor configurations and relative localization algorithms. 

Consider the relative localization of a target sensor whose real position is at $\boldsymbol{p}_T$ and its estimate is $\boldsymbol{\hat{p}}_T$.
By getting the derivative of the distance vector function $\boldsymbol{d}_{p}(\boldsymbol{p}_T)$ and taking the first-order Taylor expansion, 
we obtain $\Delta\boldsymbol{d}_{p}=H_{\boldsymbol{p}_T}\Delta\boldsymbol{p}_T$, where 
\begin{equation*}
\begin{aligned}
H_{\boldsymbol{p}_T}&=\nabla_{\boldsymbol{p}_T}\boldsymbol{d}_{p} \\
&=\begin{bmatrix}-\displaystyle\frac{x_{A_1T}-x_T}{d_{A_1T}} & -\displaystyle\frac{y_{A_1T}-y_T}{d_{A_1T}} & -\displaystyle\frac{z_{A_1T}-z_T}{d_{A_1T}}\\ \vdots & \vdots \\ -\displaystyle\frac{x_{A_kT}-x_T}{d_{A_kT}} & -\displaystyle\frac{y_{A_kT}-y_T}{d_{A_kT}} &-\displaystyle\frac{z_{A_kT}-z_T}{d_{A_kT}}\end{bmatrix}
\end{aligned}
\end{equation*} 
represents the Jacobian of $\boldsymbol{d}_{p}=\boldsymbol{d}_{p}(\boldsymbol{p}_T)$. The least squares solution of $\Delta\boldsymbol{p}_T$ is $$\Delta\boldsymbol{p}_T=(H_{\boldsymbol{p}_T}^{\top}H_{\boldsymbol{p}_T})^{-1}H_{\boldsymbol{p}_T}^{\top}\Delta\boldsymbol{d}_{p}.$$ Taking the covariance matrix of $\boldsymbol{p}_T$, we can get
$$\mathrm{Cov}(\boldsymbol{p}_T)= E[\Delta\boldsymbol{p}_T\cdot(\Delta\boldsymbol{p}_T)^{\top}]=(H_{\boldsymbol{p}_T}^{\top}H_{\boldsymbol{p}_T})^{-1}\sigma_{0}^{2},$$
where $E[\Delta\boldsymbol{d}_{p}\Delta\boldsymbol{d}_{p}^{\top}]=\mathrm{Cov}(\boldsymbol{d}_{p})=I\cdot\sigma_{0}^{2}$ is used since the measurement noise is i.i.d. Gaussian. 
The GDOP to localize a target sensor $\boldsymbol{p}_T$ is 
\begin{equation}
\mathrm{GDOP}_{\boldsymbol{p}_T}=\sqrt{\mathrm{trace}((H_{\boldsymbol{p}_T}^{\top}H_{\boldsymbol{p}_T})^{-1})}.
\label{eq:GDOP_p_T_3D}
\end{equation}
It can be proved that the best achievable RMSE of the estimator is $\sigma_{0}\cdot\mathrm{GDOP}_{\boldsymbol{p}_T}$, which is the square root of the CRLB. 

Consider the relative localization of a target agent, by getting the derivative of the distance vector function $\boldsymbol{d}_{B}(\boldsymbol{\beta}_B)$ and taking the first-order Taylor expansion,
we obtain $$\Delta\boldsymbol{d}_{B}=H_{\boldsymbol{\beta}_B}\Delta\boldsymbol{\beta}_B,$$ 
where $$H_{\boldsymbol{\beta}_B}=\nabla_{\boldsymbol{\beta}_B}\boldsymbol{d}_{B}$$
represents the Jacobian of $\boldsymbol{d}_{B}=\boldsymbol{d}_{B}(\boldsymbol{\beta}_B)$. 

Thus we have the least squares solution:
\begin{equation}
    \Delta\boldsymbol{\beta}_B=(H^{\top}_{\boldsymbol{\beta}_B}H_{\boldsymbol{\beta}_B})^{-1}H^{\top}_{\boldsymbol{\beta}_B}\Delta\boldsymbol{d}_B
\end{equation}
Taking the covariance matrix of $\boldsymbol{p}_B$, $\gamma_B$, $\theta_B$, and $\psi_B$, we can get
$$\mathrm{Cov}(\boldsymbol{\beta}_B)= E[\Delta\boldsymbol{\beta}_B\cdot\Delta\boldsymbol{\beta}_B^{\top}]=(H_{\boldsymbol{\beta}_B}^{\top}H_{\boldsymbol{\beta}_B})^{-1}\sigma_{0}^{2},$$ 
where $E[\Delta\boldsymbol{d}_{B}\Delta\boldsymbol{d}_{B}^{\top}]=\mathrm{Cov}(\boldsymbol{d}_{B})=I\cdot\sigma_{0}^{2}$ is used since the measurement noise is i.i.d. Gaussian. 
Denote $$(H_{\boldsymbol{\beta}_B}^{\top}H_{\boldsymbol{\beta}_B})^{-1}=\begin{bmatrix}
    D_{11} & D_{12} & \cdots & D_{16} \\ 
    D_{21} & D_{22} & \cdots & D_{26} \\
    \vdots & \vdots & \ddots & \vdots \\
    D_{61} & D_{62} & \cdots & D_{66} 
\end{bmatrix},$$ then the GDOP here is the trace of $(H_{\boldsymbol{\beta}_B}^{\top}H_{\boldsymbol{\beta}_B})^{-1}$ and we can also respectively get the position and pose GDOP of the target agent-$B$ by 
\begin{equation}
\mathrm{GDOP}_{\boldsymbol{p}_B}=D_{11}+D_{22}+D_{33},
\label{eq:3D_GDOP_position}
\end{equation}
\begin{equation}
    \mathrm{GDOP}_{\gamma_B}=D_{44},
    \label{eq:3D_GDOP_gamma}
\end{equation}
\begin{equation}
    \mathrm{GDOP}_{\theta_B}=D_{55},
    \label{eq:3D_GDOP_theta}
\end{equation}
\begin{equation}
    \mathrm{GDOP}_{\psi_B}=D_{66}.
    \label{eq:3D_GDOP_psi}
\end{equation}
The CRLB and the best theoretical RMSE can also be obtained from the GDOP. 

\subsection{Sensor Configuration} 
\begin{figure}[!htbp]
        \centering
        \includegraphics[width=0.4\textwidth]{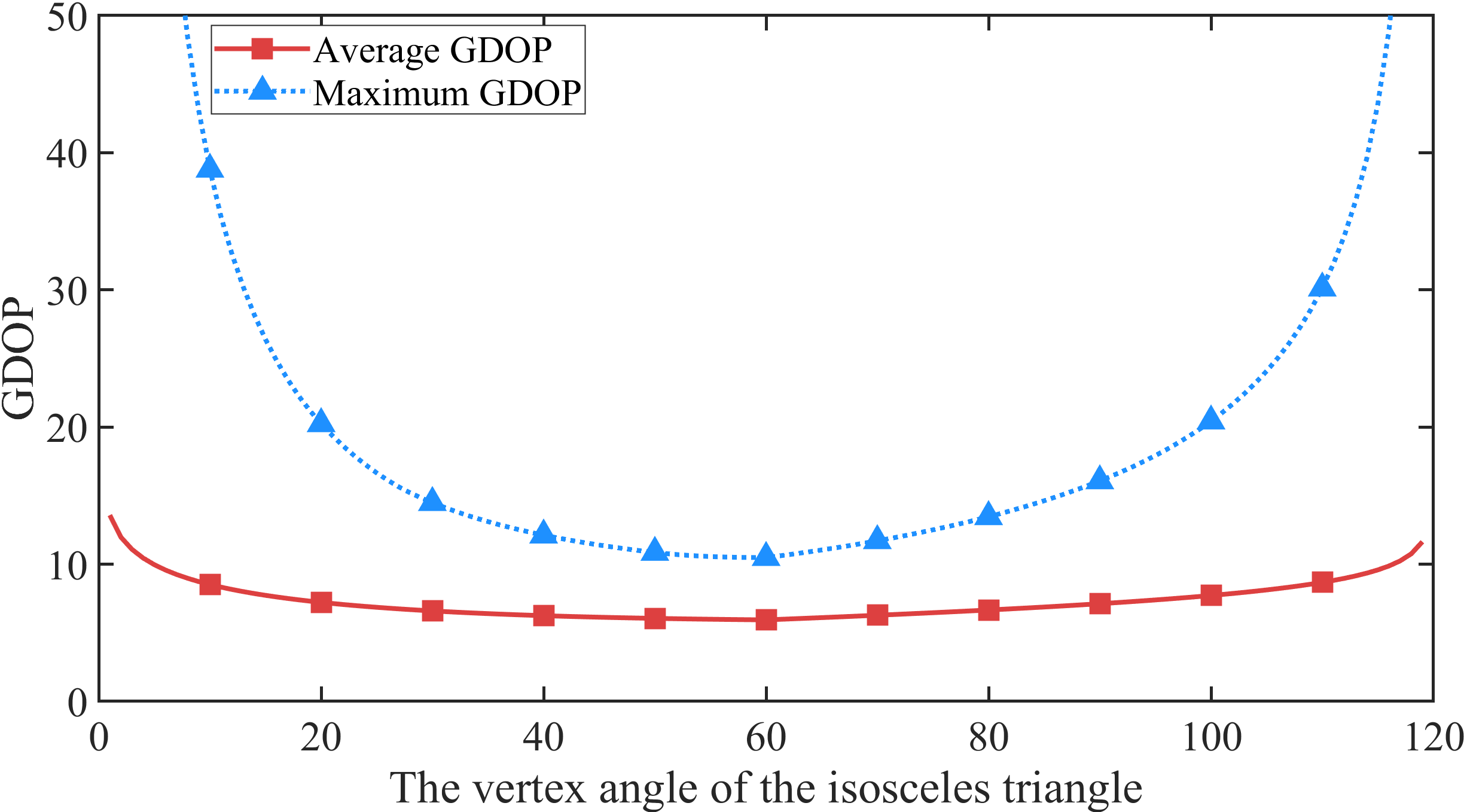}
        \caption{Comparison of the maximum and average GDOP of sensors configured on the vertexes of a tetrahedron with at least five sides equal to \SI{1}{\m} within a spherical cone with its radial distance from \SI{80}{\cm} to \SI{500}{\cm}, polar angle from \SI{0}{\degree} to \SI{180}{\degree}, and azimuthal angle from \SI{30}{\degree} to \SI{90}{\degree} in a spherical coordinate system. The vertex angle of the isosceles triangle on the side varies from $\SI{1}{\degree}$ to $\SI{119}{\degree}$. }
        \label{fig:GDOP_IsoTriangleGraph_Vertex1to119_3D}
\end{figure}

Here the GDOP, derived in detail in Section \ref{subsect:GDOP_and_CRLB_3D}, is used as a performance metric to evaluate different sensor configuration strategies. In our analysis of sensor layout optimization, we evaluated various sensor arrangements positioned on a tetrahedron's vertices, where at least five edges measured \SI{1}{\m}. We examined configurations where the vertex angle of the isosceles triangular face ranged from \SI{1}{\degree} to \SI{119}{\degree}. Due to the symmetry, the GDOP was assessed within a spherical cone, with distances varying from \SI{80}{\cm} to \SI{500}{\cm}, polar angles spanning \SI{0}{\degree} to \SI{180}{\degree}, and azimuthal angles from \SI{30}{\degree} to \SI{90}{\degree} in spherical coordinates. Both maximum and average GDOP values were computed and are shown in Fig.~\ref{fig:GDOP_IsoTriangleGraph_Vertex1to119_3D}. 

Results show that among all tested configurations, arranging sensors on the vertexes of a regular tetrahedron pattern yields superior performance. This configuration achieves both lower GDOP values and more uniform GDOP distribution, indicating superior measurement accuracy and more consistent precision across the 3D space. 



\section{Algorithms}
\label{sect:Algorithms_3D}
In this section, two state-of-the-art methods, the TT and the Fro-CVX, are first introduced. 

TT is the most commonly used range-measurement localization method with low computational complexity, and Fro-CVX transforms the localization problem to a convex optimization problem to solve with mature methods. 
Here we establish TT as our computational efficiency reference and Fro-CVX as our accuracy standard. While TT offers computational efficiency but lacks precision, Fro-CVX provides high accuracy but demands substantial computational resources, neither fully meets our low-cost application requirements. To address these limitations, we developed two novel approaches: the EDM-based trilateration (EDMT) and the maximum likelihood estimation (MLE) methods. These proposed solutions achieve the desired accuracy while maintaining computational efficiency suitable for embedded system implementation. We present both theoretical complexity analysis and empirical execution time measurements for all methods under consideration.


\subsection{The Traditional Trilateration (TT)}
\label{subsect:TT_3D}
Being the most commonly used localization method, we can extend the TT method from 2D to 3D scenarios with simple adjustment\cite{cheok_uwb_2010,thomas_revisiting_2005,li_novel_2017}.

To localize a sensor, we can estimate the position $\boldsymbol{p}_T$ by solving a linear equation $\boldsymbol{n}=J\boldsymbol{\hat{p}}_T$ derived from the Pythagorean theorem, where 
$$J=2\begin{bmatrix}
x_{A_2}-x_{A_1} & y_{A_2}-y_{A_1} & z_{A_2}-z_{A_1}\\
x_{A_3}-x_{A_2} & y_{A_3}-y_{A_2} & z_{A_3}-z_{A_2}\\
x_{A_4}-x_{A_3} & y_{A_4}-y_{A_3} & z_{A_4}-z_{A_3}\\
x_{A_1}-x_{A_4} & y_{A_1}-y_{A_4} & z_{A_1}-z_{A_4}
\end{bmatrix},$$
$\boldsymbol{n}=$\begin{footnotesize}$\begin{bmatrix}
(x_{A_2}^2-x_{A_1}^2)+(y_{A_2}^2-y_{A_1}^2)+(z_{A_2}^2-z_{A_1}^2)-(\hat{d}_{A_2T}^2-\hat{d}_{A_1T}^2) \\
(x_{A_3}^2-x_{A_2}^2)+(y_{A_3}^2-y_{A_2}^2)+(z_{A_3}^2-z_{A_2}^2)-(\hat{d}_{A_3T}^2-\hat{d}_{A_2T}^2) \\
(x_{A_4}^2-x_{A_3}^2)+(y_{A_4}^2-y_{A_3}^2)+(z_{A_4}^2-z_{A_3}^2)-(\hat{d}_{A_4T}^2-\hat{d}_{A_3T}^2) \\ 
(x_{A_1}^2-x_{A_4}^2)+(y_{A_1}^2-y_{A_4}^2)+(z_{A_1}^2-z_{A_4}^2)-(\hat{d}_{A_1T}^2-\hat{d}_{A_4T}^2)
\end{bmatrix}.$\end{footnotesize}\\ 
Then the least squares solution of the estimated target sensor position $\boldsymbol{p}_{T}$ is
\begin{equation}
\boldsymbol{\hat{p}}_T=(J^{\top}J)^{-1}J^{\top}\boldsymbol{n}
\label{eq:TT_Case_I_3D}
\end{equation} 

To localize an agent, we first apply the same method to respectively get the position estimate of the sensor $\boldsymbol{\hat{p}}_{B_i}$ on target agent $B$. Then the estimated target agent position $\boldsymbol{p}_{B}$ is 
\begin{equation}
\boldsymbol{\hat{p}}_{B}=\frac{1}{4}(\boldsymbol{\hat{p}}_{B_1}+\boldsymbol{\hat{p}}_{B_2}+\boldsymbol{\hat{p}}_{B_3}+\boldsymbol{\hat{p}}_{B_4}).
\label{eq:TT_q_B_3D}
\end{equation}
The orientation angles (yaw, pitch, and roll) can be determined through a systematic procedure. Initially, we derive the optimal rotation matrix $\mathbf{R}(\gamma,\theta,\psi)$ by solving an orthogonal Procrustes optimization problem.
\begin{equation}
(\hat{\gamma}_{B},\hat{\theta}_{B},\hat{\psi}_{B})=\arg\min\limits_{\gamma,\theta,\psi}\|(\hat{P}_{B}-\boldsymbol{\hat{p}}_{B}\mathbf{1}^{\top})-\mathbf{R}(\gamma,\theta,\psi)P_A\|_F,
\label{eq:TT_hat_ypr_B_3D}
\end{equation}
where $\hat{P}_{B}=\begin{bmatrix}
    \boldsymbol{\hat{p}}_{B_1} & \boldsymbol{\hat{p}}_{B_2} & \boldsymbol{\hat{p}}_{B_3} & \boldsymbol{\hat{p}}_{B_4}\end{bmatrix}, P_A=\begin{bmatrix}
    {\boldsymbol{p}}_{1} & {\boldsymbol{p}}_{2} & {\boldsymbol{p}}_{3} & {\boldsymbol{p}}_{4} \end{bmatrix}$, 
and $\mathbf{R}(\gamma,\theta,\psi)$ is the rotation matrix defined in Eq. (\ref{eq:YawPitchRoll_RotationMatrix}). We refer to the orthogonal Procrustes problem and the Eq. (\ref{eq:Get_YawPitchRollfromR}) to address the optimization problem in Eq. (\ref{eq:TT_hat_ypr_B_3D}). In the standard orthogonal Procrustes problem, we solve the optimization problem where $Q$ is orthogonal by \begin{equation}
\arg\min_{Q}\|X-QY\|=VU^{\top},
\label{eq:ProcrustesSolution}
\end{equation} where $U$, $V$ are from the singular value decomposition (SVD) $YX^{\top} = USV^{\top}$ given $X$ and $Y$ \cite{gao_TIM_2025_InfrastructureFreeRelativeLocalization}.  
Let $X=\hat{P}_{B}-\boldsymbol{\hat{p}}_{B}\mathbf{1}^{\top}$ and $Y=P_A$, we get 
$\mathbf{R}(\gamma,\theta,\psi)=V_{R}U^{\top}_{R}$, where $V_{R},U_{R}$ are from the SVD $P_A(\hat{P}_{B}-\boldsymbol{\hat{p}}_{B}\mathbf{1}^{\top})^{\top}=U_{R}SV^{\top}_{R}$.
Then the target yaw, pitch, and roll angle estimates $\hat{\gamma}_B,\hat{\theta}_B$, and $\hat{\psi}_B$ can be derived from Eq. (\ref{eq:Get_YawPitchRollfromR}). 

\subsection{Frobenius-norm-based Estimation (Fro-CVX)}
\label{subsect:Fro-CVX_3D}
In addition to the TT algorithm, Fro-CVX represents another widely-adopted approach for sensor network localization. This method transforms the localization problem into a convex optimization framework based on the Frobenius norm \cite{cao_sensor_2006,so_theory_2007,cao_formation_2011}. 

Consider $n$ points $\boldsymbol{p}_i\in\mathbb{R}^3,i=1,\ldots,n$. Denote the position matrix $P\in\mathbb{R}^{3\times n}$ of the points as 
\begin{equation}
P=\begin{bmatrix} \boldsymbol{p}_1 & \boldsymbol{p}_2 & \cdots & \boldsymbol{p}_n \end{bmatrix},
\label{eq:PositionMatrix_3D}
\end{equation} the estimated position matrix $\hat{Q}\in\mathbb{R}^{3\times n}$ as 
\begin{equation}
\hat{P}=\begin{bmatrix} \boldsymbol{\hat{p}}_1 & \boldsymbol{\hat{p}}_2 & \cdots & \boldsymbol{\hat{p}}_n \end{bmatrix},
\label{eq:EstimatedPositionMatrix_3D}
\end{equation} and the corresponding EDM $E\in\mathbb{R}^{n\times n}$ as 
\begin{equation}
[E_{P}]_{(i,j)}=\|\boldsymbol{p}_i-\boldsymbol{p}_j\|^2=\boldsymbol{p}_{i}^{\top}\boldsymbol{p}_i-2\boldsymbol{p}_{i}^{\top}\boldsymbol{p}_{j}+\boldsymbol{p}_{j}^{\top}\boldsymbol{p}_j,
\label{eq:EDM_3D}
\end{equation} where $[E_{P}]_{(i,j)}$ is the $(i,j)$-th element of $E_{P}$ \cite{dattorro_convex_2019}. 

To localize a sensor $T$ with four sensors on agent $A$, we have the measured distance matrix and the real EDM of the target in the form 
        $$\hat{E}_{P_T}=\begin{bmatrix}
          F_{P} & \hat{\boldsymbol{e}}_{Q} \\ \hat{\boldsymbol{e}}_{Q}^{\top} & 0
        \end{bmatrix},
        E_{P_T}(\boldsymbol{p}_T)=\begin{bmatrix}
          F_{P} & \boldsymbol{e}_{Q}(\boldsymbol{p}_T) \\ \boldsymbol{e}_{Q}^{\top}(\boldsymbol{p}_T) & 0
        \end{bmatrix},$$
        where $$\hat{\boldsymbol{e}}_{Q}=\begin{bmatrix}
        \hat{d}_{A_1T}^2 & \hat{d}_{A_2T}^2 & \hat{d}_{A_3T}^2 & \hat{d}_{A_4T}^2 \end{bmatrix}^{\top},$$
        $${\boldsymbol{e}_{Q}}(\boldsymbol{p}_T)=\begin{bmatrix}
        {d}_{A_1T}^2 (\boldsymbol{p}_T) & d_{A_2T}^2 (\boldsymbol{p}_T) & d_{A_3T}^2 (\boldsymbol{p}_T) & d_{A_4T}^2 (\boldsymbol{p}_T) \end{bmatrix}^{\top}$$ 
        are respectively from the measurement and from the real distance function with $d_{A_iT}(\boldsymbol{p}_T)$ defined in Eq. (\ref{eq:d_AiT_3D}). $F_{P}$ is known \textit{a priori} from the sensor configuration as  
    \begin{equation}
F_{P}=\begin{bmatrix}0&d_{12}^2&d_{13}^2&d_{14}^2\\d_{21}^2&0&d_{23}^2&d_{24}^2\\d_{31}^2&d_{32}^2&0&d_{34}^2\\d_{41}^2&d_{42}^2&d_{43}^2&0\end{bmatrix},
    \label{eq:F_P}
    \end{equation}

        with $d_{ij}=\|\boldsymbol{p}_i-\boldsymbol{p}_j\|,i,j=1,\ldots,4$.

To localize an agent $B$, we have the measured distance matrix and the real EDM of the target agent in the form 
    $$\hat{E}_{P_B}=\begin{bmatrix} F_{P} & \hat{E}_{P,AB} \\ \hat{E}_{P,AB}^{\top} & F_{P} \end{bmatrix},$$
    and 
    $$E_{P_B}(\boldsymbol{\beta}_B)=\begin{bmatrix} F_{P} & E_{P,AB}(\boldsymbol{\beta}_B) \\ E_{P,AB}^{\top}(\boldsymbol{\beta}_B) & F_{P} \end{bmatrix},$$
    where the (i,j)-th element of $\hat{E}_{P,AB}$ is $\hat{E}_{P,AB}(i,j)=\hat{d}_{A_iB_j}^2$ from the measurement and the (i,j)-th element of $E_{P,AB}(\boldsymbol{\beta}_B)$ is $E_{P,AB}(\boldsymbol{\beta}_B)(i,j)=d_{A_iB_j}^2 (\boldsymbol{\beta}_B)$ from the real distance function with $d_{A_{i}B_{j}}(\boldsymbol{\beta}_{{B}})=\|\boldsymbol{p}_{A_i}-\boldsymbol{p}_{B_j}(\boldsymbol{\beta}_{{B}})\|,i,j=1,2,3,4$. $F_{P}$ is defined in Eq. (\ref{eq:F_P}) and $\boldsymbol{p}_{B_j}(\boldsymbol{\beta}_{{B}})$ defined in Eq. (\ref{eq:q_Bj_3D}).

This method aims to find the $\boldsymbol{\hat{p}}_T$ and $\boldsymbol{\hat{\beta}}_B$ to minimize $\|\hat{E}_{P_T}-E_{P_T}(\boldsymbol{p}_T)\|_{F}$ and $\|\hat{E}_{P_B}-E_{P_B}(\boldsymbol{\beta}_T)\|_{F}$, respectively. 
Then we can get the target state by solving the optimization problems: 
\begin{equation}\boldsymbol{\hat{p}}_T=
\arg\min\limits_{{\boldsymbol{p}}_T} \|\hat{E}_{P_T}-E_{P_T}(\boldsymbol{p}_T)\|_F,
\label{eq:Fro-CVX_Case_I_3D}
\end{equation}
\begin{equation}
\boldsymbol{\hat{\beta}}_B=\arg\min\limits_{\boldsymbol{\beta}_B} \|\hat{E}_{P_B}-E_{P_B}(\boldsymbol{\beta}_B)\|_F.
\label{eq:Fro-CVX_Case_II_3D}
\end{equation}

Obviously, the optimal solutions in Eq. (\ref{eq:Fro-CVX_Case_I_3D}) and Eq. (\ref{eq:Fro-CVX_Case_II_3D}) are, respectively, equivalent to Eq. (\ref{eq:Fro-CVX_Case_I_3D_equi}) and Eq. (\ref{eq:Fro-CVX_Case_II_3D_equi}):
\begin{equation}
 \boldsymbol{\hat{p}}_T=\arg\min\limits_{{\boldsymbol{p}}_T} \sum\limits_{i=1}^{4}\left(\hat{d}_{A_iT}^2-d_{A_iT}^2({\boldsymbol{p}}_{{T}})\right)^2,
 \label{eq:Fro-CVX_Case_I_3D_equi}
\end{equation}
\begin{equation}
\boldsymbol{\hat{\beta}}_B=\arg\min\limits_{\boldsymbol{\beta}_B}\sum\limits_{i,j\in\{1,2,3,4\}}\left(\hat{d}_{A_iB_j}^2-d_{A_iB_j}^2(\boldsymbol{\beta}_{{B}})\right)^2.
 \label{eq:Fro-CVX_Case_II_3D_equi}
\end{equation}
 
The problems can be efficiently solved using numerical techniques, including Newton's iterative method or readily available commercial convex optimization software. 

\subsection{The EDM-based Trilateration (EDMT)}
\label{subsect:EDMT_3D}
EDM-based localization techniques have been extensively studied and applied across diverse fields, from molecular conformation analysis to sensor network positioning \cite{chu_least_2008,gao_euclidean_2019}. Our proposed EDMT method extends beyond traditional target sensor localization to include agent localization capabilities.
Through the application of EDM theory, we achieve a notable reduction in computational complexity.

For a set of $n$ points $\boldsymbol{p}_i\in\mathbb{R}^3,i=1,\ldots,n$, we define the position matrix $P$ and its corresponding EDM $E_{P}$ according to Eq. (\ref{eq:PositionMatrix_3D}) and Eq. (\ref{eq:EDM_3D}). Let $G_{P}=P^{\top}P$, where the $(i,j)$-th element of $G_{P}$ is given by $[G_{P}]{(i,j)}=\boldsymbol{p}_i^{\top}\boldsymbol{p}_j$. The relationship between $E_{P}$ and $G_{P}$ can be expressed by the following equation:
\begin{equation}\label{eq:G_EDM_3D}
E_{P}=\mathbf{1}_{n}\mathrm{diag}(G_{P})^{\top}-2G_{P}+\mathrm{diag}(G_{P})\mathbf{1}_{n}^{\top},
\end{equation} 
where $\mathrm{diag}(G_{P})$ represents a column vector containing the diagonal elements of $G$ in sequential order. For clarity, we define the following sets:
\begin{itemize}
        \item $\mathbb{R}^{n\times n}_{+}$: the set of all $n\times n$ real matrices with nonnegative elements.
        \item $\mathbb{S}^{n}_{h}$: the set of hollow symmetric $n\times n$ matrices (symmetric matrices with zero diagonal elements).
        \item $\mathbb{EDM}^{n}$: the set of $n\times n$ EDM matrices.
\end{itemize}
As noted in \cite{dattorro_convex_2019}, $\mathbb{EDM}^{n}$ is a subset of the intersection of $\mathbb{R}^{n\times n}_{+}$ and $\mathbb{S}^{n}_{h}$, i.e., $\mathbb{EDM}^{n}\subset (\mathbb{R}^{n\times n}_{+}\cap\mathbb{S}^{n}_{h})$.

In practical applications, range measurements between sensors contain measurement error. Consequently, the measured distance matrices $\hat{E}_{P_T}$ and $\hat{E}_{P_B}$ belong to $(\mathbb{R}^{n\times n}_{+}\cap\mathbb{S}^{n}_{h})$ but may not qualify as true EDM matrices. To estimate the target state for a sensor and an agent, we employ the following systematic procedures.
\subsubsection{Distinguishing an EDM}
The initial step involves determining whether a given measured distance matrix $\hat{E}_{P}$ constitutes an EDM, specifically whether there exists a corresponding position matrix $P$ that generates $\hat{E}_{P}$ as its EDM. Theorem \ref{thm:EDM_NSConditions} \cite{menger_new_1931,schoenberg_remarks_1935} outlines the necessary and sufficient conditions for identifying a valid EDM.
\begin{theorem}[Necessary and sufficient conditions of an EDM]
\label{thm:EDM_NSConditions}
A distance matrix $E\in(\mathbb{R}^{n\times n}_{+}\cap\mathbb{S}^{n}_{h})$ is an EDM if and only if $-V^{\top}EV$ is positive semi-definite, where the range of $V$ equals to the orthogonal complement of $\mathbf{1}$, i.e., $\mathcal{R}(V)=\mathbf{1}^{\perp}$. 

The commonly used matrix $V$ includes the Schoenberg auxiliary matrix $V_{S}$ \cite{schoenberg_remarks_1935} and the geometric centering matrix $V_{c}$, where  \begin{equation}
V_{S}=\frac{1}{\sqrt{2}}\begin{bmatrix}
-\mathbf{1}_{n-1}^{\top}\\
I_{n-1} 
\end{bmatrix},V_{c}=I-\frac{1}{n}\mathbf{1}_n\mathbf{1}_n^{\top}.
\label{eq:AuxiliaryMatrix_V}
\end{equation}
\end{theorem}
\subsubsection{Finding the best-fit EDM}

For cases where $\hat{E}_{P}$ is not an EDM, or where $\hat{E}_{P}$ is an EDM but lacks a corresponding position matrix $P$ with $\mathrm{rank}(P)=3$, we must find an EDM $E_{P}\in\mathbb{EDM}^{n}$ with $\mathrm{rank}(V^{\top}E_{P}V)=3$ that best approximates $\hat{E}$ under a suitable norm to estimate the position matrix $P$. Using the Frobenius matrix norm leads to an optimization problem which may be non-convex due to the rank constraint:

\begin{align}
\min\limits_{E_{P}\in\mathbb{EDM}^{n}} & \|\hat{E}_{P}-E_{P}\|_{F} \notag\\
\text{s.t.}
        &\  \mathcal{R}(V)=\mathrm{ker}(\mathbf{1}_n^{\top}) \notag \\
        &\  \mathrm{rank}(V^{\top}E_{P}V)=3 \notag
\end{align}

To simplify this problem in our application, we introduce an alternative norm specifically designed for hollow symmetric matrices. Let the matrix $p$-norm be defined as $\|A\|_p\triangleq (\sum\limits_{i=1}^{n} | \lambda_{i} | ^p)^{\frac{1}{p}}$, where $\lambda_{i}$'s represent the eigenvalues of $A$. We then define a norm on $\mathbb{S}^{n}_{h}$ as $\|A\|^{(s)}_{p} \triangleq \|P_{s}AP_{s}^{\top}\|{p}$, where $P_{s}=I-\mathbf{1}_{n}s^{\top}$ with $s^{\top}\mathbf{1}_{n}=1$.

Then we can derive the optimal EDM $E_{P}$ with respect to the specific $|\cdot|^{(s)}_p$ norm from $\hat{E}_{P}$ by applying Theorem \ref{thm:closestEDM} \cite{mathar_best_1985}, and subsequently compute a solution $P$. It is important to note that the solution $P$ corresponding to $E_P$ is not unique. As shown in \cite[Chapter 5]{dattorro_convex_2019}, the EDM remains invariant under rigid transformations including rotation, reflection, and translation. Specifically, the EDM of $\mathbf{R}P+\boldsymbol{t}\mathbf{1}_{n}^{\top}$ equals the EDM of $P$, where $\mathbf{R}$ represents an orthogonal matrix and $\boldsymbol{t}\in\mathbb{R}^3$. 
\begin{theorem}[Closest EDM fitting theorem]
\label{thm:closestEDM}
Let $\hat{E}\in (\mathbb{R}^{n\times n}_{+}\cap\mathbb{S}^{n}_{h})$ be the measured distance matrix and $s\in\mathbb{R}^n$ be such that $s^{\top}\mathbf{1}_{n}=1$. Assume the diagonal decomposition $P_{s}(-\hat{E})P_{s}=T\mathrm{diag}(\lambda_{1},\ldots,\lambda_{n})T^{\top}$, where $P_{s}=I_{n}-\mathbf{1}_{n}s^{\top}$, $T$ is orthogonal, and $\lambda_{1}\geq\ldots\geq\lambda_{n}$. Denote $\lambda_{i}^+=\max\{\lambda_{i},0\}$. Then let $B=T\mathrm{diag}(\lambda_{1}^+,\ldots,\lambda_{r}^+,0,\ldots,0)T^{\top}$ and $E=-(B-\boldsymbol{b}\mathbf{1}_{n}^{\top}-\mathbf{1}_{n}\boldsymbol{b}^{\top})$, where $\boldsymbol{b}=\frac{1}{2}\mathrm{diag}(B)$. Then $E\in\mathbb{EDM}^{n}$ and $\|\hat{E}-E\|^{(s)}_p\leq\|\hat{E}-A\|^{(s)}_p$ for all $p\geq 1$, for all $A\in\mathbb{EDM}^{n}$ with $\mathrm{rank}(V_{c}AV_{c})\leq r<n$, where $V_{c}$ is the geometric centering matrix in Eq. (\ref{eq:AuxiliaryMatrix_V}). 
\end{theorem}
\subsubsection{Solving a special position matrix}

Given the non-uniqueness of position matrix solutions for an EDM $E_{P}\in \mathbb{EDM}^{n}$, our approach involves two steps:
\begin{itemize}
        \item Find a special solution $P^{\star}\in\mathbb{R}^{3\times n}$.
        \item Using known onboard sensor positions $\boldsymbol{p}_i, i=1,2,3,4$, determine an orthogonal matrix $N$ and a vector $\boldsymbol{t}$ to obtain $P=NP^{\star}+\boldsymbol{t}\mathbf{1}_{n}^{\top}$, where $P$ has its first four columns aligned with $\boldsymbol{p}_i$, $i=1,2,3,4$.
\end{itemize}

Let $P^{0}$ denote the special solution where the first column $\boldsymbol{p}^{0}_{1}=\mathbf{0}$. For this $P^{0}$, we have:
$$[E_{P}]_{(i,1)}=\|\boldsymbol{p}^{0}_{i}-\boldsymbol{p}^{0}_{1}\|^2=\|\boldsymbol{p}^{0}_{i}\|^2,i=1,2,\ldots,n.$$

Let $G^{0}_{P}=(P^{0})^{\top} P^{0}$ and denote the first column of $E_{P}$ as $\boldsymbol{e}_{P,1}$. Since $[G^{0}_{P}]_{(i,i)}=\|\boldsymbol{p}^{0}_{i}\|^2=[E_{P}]_{(i,1)}$, we obtain $\boldsymbol{e}_{P,1}=\mathrm{diag}(G_{P}^0)$. From Eq. (\ref{eq:G_EDM_3D}), we can derive:

\begin{equation}
 G^{0}_{P}=(P^{0})^{\top} P^{0}=-\frac{1}{2}(E_{P}-\mathbf{1}\boldsymbol{e}_{P,1}^{\top}-\boldsymbol{e}_{P,1}\mathbf{1}^{\top}).
 \label{eq:G^0_3D}
 \end{equation} 

Given that $G^{0}_{P}$ is positive semi-definite, we can solve for $P^{0}$ using eigenvalue decomposition \cite{strang_linear_2019}:

\begin{equation}
\begin{aligned}
G^{0}_{P} &= \begin{bmatrix} U_{1}^{0\top} & U_{2}^{0\top}\end{bmatrix} \mathrm{diag}(\Lambda_{3},0_{n-3}) \begin{bmatrix} U_{1}^0 \\ U_{2}^0\end{bmatrix} \\
&= \begin{bmatrix} \Lambda_{3}^{\frac{1}{2}}U_{1}^0\end{bmatrix}^{\top} \begin{bmatrix} \Lambda_{3}^{\frac{1}{2}}U_{1}^0\end{bmatrix}=(P^{0})^{\top}P^{0},
\end{aligned}
\label{eq:P^{0}_3D}
\end{equation}
where $P^{0}=\Lambda_{3}^{\frac{1}{2}}U_1^0$ represents the special solution with the first column being a zero vector. Alternative methods such as Cholesky decomposition or $LDL^{\top}$ decomposition can also be employed.
\subsubsection{Getting best fit rigid transform}
Via the method in \cite{eggert_estimating_1997}, we can compute the optimal orthogonal matrix \( \mathbf{R} \) and translation vector \( \boldsymbol{t} \) for target estimation. The translation vector is determined through centroid alignment:

\begin{equation}
\boldsymbol{t} = \frac{1}{4}\sum\limits_{i=1}^{4}(\boldsymbol{p}^{0}_{i} - \boldsymbol{p}_{i}).
\label{eq:translation_vector_t_3D}
\end{equation}
The translated solution \( P^{1} \) becomes
\[
P^{1} = P^{0} - \boldsymbol{t}\mathbf{1}_{n}^{\top}=\begin{bmatrix}
        \boldsymbol{p}^{1}_{1} & \cdots & \boldsymbol{p}^{1}_{n}
\end{bmatrix}
\]
yielding translated coordinates where the centroid of \( \boldsymbol{p}^{1}_{1} \), \( \boldsymbol{p}^{1}_{2} \), \( \boldsymbol{p}^{1}_{3} \), and \( \boldsymbol{p}^{1}_{4} \) coincides with the origin. The rotation matrix \( \mathbf{R}_1 \) is subsequently derived from the Procrustes optimization:

\[
\mathbf{R}_1 = \arg\min\limits_{\mathbf{R}} \left\| \begin{bmatrix}
\boldsymbol{p}^{1}_{1} & \boldsymbol{p}^{1}_{2} & \boldsymbol{p}^{1}_{3} & \boldsymbol{p}^{1}_{4}
\end{bmatrix} - \mathbf{R}P_A \right\|_F
\]

Following the solution methodology in Eq. (\ref{eq:ProcrustesSolution}), we obtain the position matrix estimate:
\[
\hat{P} = \mathbf{R}_1P^{1} = \mathbf{R}_1(P^{0} - \boldsymbol{t}\mathbf{1}_{n}^{\top})
\]
The target sensor position is finally estimated as:
\begin{equation}
\boldsymbol{\hat{p}}_{T} = \boldsymbol{\hat{p}}_{5}.
\label{eq:q_T}
\end{equation}
The target agent localization can be performed through two distinct methodologies: sensor-wise localization (EDMT-individually) or integrated sensor network localization (EDMT-jointly). For EDMT-individually, the position estimates $\boldsymbol{\hat{p}}_{B_i},i=1,2,3,4$ are obtained by replicating the procedure to localize a sensor.

For EDMT-jointly, the system dimensionality expands to $n=8$, with the sensor positions on the target agent corresponding to $\boldsymbol{\hat{p}}_{B_1} = \boldsymbol{\hat{p}}_{5}$, $\boldsymbol{\hat{p}}_{B_2} = \boldsymbol{\hat{p}}_{6}$, $\boldsymbol{\hat{p}}_{B_3} = \boldsymbol{\hat{p}}_{7}$, $\boldsymbol{\hat{p}}_{B_4} = \boldsymbol{\hat{p}}_{8}$.
The complete positional and angular state estimates are subsequently derived through the solutions of Eq. (\ref{eq:TT_q_B_3D}) for coordinates and Eq. (\ref{eq:TT_hat_ypr_B_3D}) for attitude angles.
The proposed methodology is systematically formalized through two dedicated algorithms:
\begin{itemize}
    \item Algorithm \ref{alg:EDM_basedPoint_3D}: EDMT implementation for individual sensor localization.
    \item Algorithm \ref{alg:EDM_basedAgent_3D}: EDMT-jointly framework for integrated agent localization.
\end{itemize}

\begin{algorithm}
\caption{3D EDMT for sensor localization}
\label{alg:EDM_basedPoint_3D}
\begin{algorithmic}
\Require {Sensor configuration $\boldsymbol{p}_{A_i}=\boldsymbol{p}_{i}$, measured distances $\hat{d}_{A_iT}$, where $i=1,\ldots,4$ }
\Ensure {The estimate $\boldsymbol{\hat{p}}_{T}$ of ${\boldsymbol{p}}_{T}$}
\State $[\hat{E}]_{(k,k)} \gets 0,k=1,\ldots,5$
\State $[\hat{E}]_{(i,j)} \gets \|\boldsymbol{p}_i-\boldsymbol{p}_j\|^2,i,j=1,2,3,4,i\neq j$
\State $[\hat{E}]_{(i,5)} \gets \hat{d}_{A_T}^2,i=1,\ldots,5$
\State $[\hat{E}]_{(5,j)} \gets \hat{d}_{A_jT}^2,j=1,\ldots,5$
\State $P_A\gets\begin{bmatrix} \boldsymbol{p}_{A_1} & \boldsymbol{p}_{A_2} & \boldsymbol{p}_{A_3}& \boldsymbol{p}_{A_4}\end{bmatrix}$\;

\If{$\hat{E}\notin \mathbb{EDM}^{5}$ or $\mathrm{rank}(-V^{\top}\hat{E}V)> 3$, where $V=I-\frac{1}{5}\mathbf{1}\mathbf{1}^{\top}$}
  \State {diagonal decomposition: $$-V^{\top}\hat{E}V=T\mathrm{diag}(\lambda_1,\lambda_2,\lambda_3,\lambda_4,\lambda_5)T^{\top}$$, where $T$ is orthogonal, $\lambda_1\geq\lambda_2\geq\lambda_3\geq\lambda_4\geq\lambda_5$\;
  \State $B=T\mathrm{diag}(\lambda_1^{+},\lambda_2^{+},\lambda_3^{+},0,0)T^{\top}$, where $\lambda_1^{+}= \max\{\lambda_1,0\},\lambda_2^{+}= \max\{\lambda_2,0\},\lambda_3^{+}= \max\{\lambda_3,0\}$\;
  \State $E=-(B-\boldsymbol{b}\mathbf{1}^{\top}-\mathbf{1}\boldsymbol{b}^{\top}),\boldsymbol{b}=\frac{1}{2}\mathrm{diag}(B)$\;}
\Else 
  {\ $E=\hat{E}$\;}
\EndIf
\State $G^{0}=-\frac{1}{2}(E-\mathbf{1}\boldsymbol{e}_1^{\top}-\boldsymbol{e}_1\mathbf{1}^{\top})$\; 
\State Eigenvalue decomposition of $G^{0}$: $G^{0}=\begin{bmatrix}U_1^{0\top} & U_2^{0\top}\end{bmatrix}\mathrm{diag}(\Lambda_2,0_2)\begin{bmatrix}U_1^{0\top} & U_2^{0\top}\end{bmatrix}^{\top}$\; 
\State Compute the special position matrix: $P^{0}=\Lambda_{2}^{\frac{1}{2}}U_1^0=\begin{bmatrix}\boldsymbol{p}_1^0 & \boldsymbol{p}_2^0 & \boldsymbol{p}_3^0 & \boldsymbol{p}_4^0 & \boldsymbol{p}_5^0\end{bmatrix}$\; 
\State Compute the translation vector: $\boldsymbol{t}=\frac{1}{4}\sum_{i=1}^{4}(\boldsymbol{p}_i^0-\boldsymbol{p}_i)$\; 
\State Compute the translated position matrix: $Q^1=P^{0}-\boldsymbol{t}\mathbf{1}_5^{\top}=\begin{bmatrix}\boldsymbol{p}_1^1 & \boldsymbol{p}_2^1 & \boldsymbol{p}_3^1 & \boldsymbol{p}_4^1 & \boldsymbol{p}_5^1\end{bmatrix}$\; 
\State Compute the rotation matrix: $\mathbf{R}_1=V_1U_1^{\top}$, where $U_1,V_1$ are from the SVD $P_A\begin{bmatrix}\boldsymbol{p}_1^1 & \boldsymbol{p}_2^1 & \boldsymbol{p}_3^1 & \boldsymbol{p}_4^1 \end{bmatrix}^{\top}=U_1S_1V_1^{\top}$\; 
\State Compute the target position: $\boldsymbol{\hat{p}}_{T}=\mathbf{R}_1\boldsymbol{p}^{1}_{5}$.
\end{algorithmic}
\end{algorithm}

\begin{algorithm}
\caption{3D EDMT-jointly for agent localization}
\label{alg:EDM_basedAgent_3D}
\begin{algorithmic}
\Require {Sensor configuration $\boldsymbol{p}_{A_i}=\boldsymbol{p}_{i},i=1,\ldots,4$, measured distances $\hat{d}_{A_iB_j},i,j=1,\ldots,4$ }
\Ensure {The estimate $\boldsymbol{\hat{p}}_{B_j}$ of ${\boldsymbol{p}}_{B_j},j=1,\ldots,4$, the estimate $\boldsymbol{\hat{p}}_{B}$ of centroid ${\boldsymbol{p}}_{B}$, and the estimate $\hat{\theta}_{B}$ of ${\theta}_{B}$}
\State $[\hat{E}]_{(k,k)} \gets 0,k=1,\ldots,8$
\State $[\hat{E}]_{(i,j)}\gets\|\boldsymbol{p}_i-\boldsymbol{p}_j\|^2,i,j=1,\ldots,4$ 
\State $[\hat{E}]_{(4+i,4+j)} \gets \|\boldsymbol{p}_i-\boldsymbol{p}_j\|^2,i,j=1,\ldots,4$
\State $[\hat{E}]_{(i,4+j)}\gets\hat{d}_{A_iB_j}^2,i,j=1,\ldots,4$ 
\State $[\hat{E}]_{(4+j,i)}\gets\hat{d}_{A_iB_j}^2,i,j=1,\ldots,4$

\State $P_A \gets \begin{bmatrix} \boldsymbol{p}_{A_1} & \boldsymbol{p}_{A_2} & \boldsymbol{p}_{A_3}\end{bmatrix}$\;

\If{$\hat{E}\notin \mathbb{EDM}^{8}$ or $\mathrm{rank}(-V^{\top}\hat{E}V)> 3$, where $V=I-\frac{1}{8}\mathbf{1}\mathbf{1}^{\top}$}
  \State {diagonal decomposition: $-V^{\top}\hat{E}V=T\mathrm{diag}(\lambda_1,\lambda_2,\ldots,\lambda_8)T^{\top}$, where $T$ is orthogonal, $\lambda_1\geq\lambda_2\geq\ldots\geq\lambda_8$\;
  \State $B=T\mathrm{diag}(\lambda_1^{+},\lambda_2^{+},\lambda_3^{+},0,0)T^{\top}$, where $\lambda_1^{+}= \max\{\lambda_1,0\},\lambda_2^{+}= \max\{\lambda_2,0\},\lambda_3^{+}= \max\{\lambda_3,0\}$\;
  \State $E=-(B-\boldsymbol{b}\mathbf{1}^{\top}-\mathbf{1}\boldsymbol{b}^{\top}),\boldsymbol{b}=\frac{1}{2}\mathrm{diag}(B)$\;}
\Else 
  {\ $E=\hat{E}$\;}
\EndIf
\State $G^{0}=-\frac{1}{2}(E-\mathbf{1}\boldsymbol{e}_1^{\top}-\boldsymbol{e}_1\mathbf{1}^{\top})$\; 
\State Eigenvalue decomposition of $G^{0}$: $G^{0}=\begin{bmatrix}U_1^{0\top} & U_2^{0\top}\end{bmatrix}\mathrm{diag}(\Lambda_2,0_4)\begin{bmatrix}U_1^{0\top} & U_2^{0\top}\end{bmatrix}^{\top}$\; 
\State Compute the special position matrix: $P^{0}=\Lambda_{3}^{\frac{1}{2}}U_1^0=\begin{bmatrix}\boldsymbol{p}_1^0 & \boldsymbol{p}_2^0 & \cdots & \boldsymbol{p}_8^0\end{bmatrix}$\; 
\State Compute the translation vector: $\boldsymbol{t}=\frac{1}{4}\sum_{i=1}^{4}(\boldsymbol{p}_i^0-\boldsymbol{p}_i)$\; 
\State Compute the translated position matrix: $Q^1=P^{0}-\boldsymbol{t}\mathbf{1}_8^{\top}=\begin{bmatrix}\boldsymbol{p}_1^1 & \boldsymbol{p}_2^1 & \cdots & \boldsymbol{p}_8^1\end{bmatrix}$\; 
\State Compute the rotation matrix: $\mathbf{R}_1=V_1U_1^{\top}$, where $U_1,V_1$ are from the SVD $P_A\begin{bmatrix}\boldsymbol{p}_1^1 & \boldsymbol{p}_2^1 & \boldsymbol{p}_3^1 \end{bmatrix}^{\top}=U_1S_1V_1^{\top}$\; 
\State Compute the target agent vertexes: $\boldsymbol{\hat{p}}_{B_1}=\mathbf{R}_1{\boldsymbol{p}}^{1}_{5}$, $\boldsymbol{\hat{p}}_{B_2}=\mathbf{R}_1{\boldsymbol{p}}^{1}_{6}$, $\boldsymbol{\hat{p}}_{B_3}=\mathbf{R}_1{\boldsymbol{p}}^{1}_{7}$, and $\boldsymbol{\hat{p}}_{B_4}=\mathbf{R}_1{\boldsymbol{p}}^{1}_{8}$, the position matrix $\hat{Q}_B=\begin{bmatrix}\boldsymbol{\hat{p}}_{B_1} & \boldsymbol{\hat{p}}_{B_2} & \boldsymbol{\hat{p}}_{B_3} & \boldsymbol{\hat{p}}_{B_4}\end{bmatrix}$, the target centroid $\boldsymbol{\hat{p}}_{B}=\frac{1}{4}\sum_{i=1}^{4}(\boldsymbol{\hat{p}}_{B_i})$\; 
\State Compute the target yaw, pitch, and roll angles: $\hat{\psi}_B=\arctan\frac{-\mathbf{R}_{12}}{\mathbf{R}_{11}},\hat{\theta}=\arctan\frac{\mathbf{R}_{13}}{\mathbf{R}_{23}^2+\mathbf{R}_{33}^2}, \hat{\gamma}=\arctan\frac{-\mathbf{R}_{23}}{\mathbf{R}_{33}}$, where $\mathbf{R}=V_2U_2^{\top}$, $V_2,U_2$ are from the SVD $P_A(\hat{Q}_B-\boldsymbol{\hat{p}}_{B}\mathbf{1}^{\top})^{\top}=U_2S_2V_2^{\top}$\;
\end{algorithmic}
\end{algorithm}

\subsection{Maximum Likelihood Estimation (MLE)}
\label{subsect:MLE_3D}
The Maximum Likelihood Estimator (MLE) is widely recognized for its asymptotic optimality. Specifically, it achieves the CRLB under conditions of high signal-to-noise ratio (SNR) and a large number of measurements \cite{kay_fundamentals_1993}. In the absence of \textit{a priori} information regarding the deterministic unknowns, MLE generally represents the optimal estimation approach. 
Based on the experimental results from static range measurements using UWB ranging sensors, the measurement error characteristics can be effectively approximated as Gaussian white noise with a standard deviation of approximately $\SI{5}{\cm}$ for ranges below $\SI{600}{\cm}$. In this subsection, we model both the sensor and agent localization measurement errors as i.i.d. random variables with a standard deviation of $\sigma_0$.

For sensor localization, given the Gaussian noise $\epsilon_i\sim \mathcal{N}(0,\sigma_{0}^2)$ for $i=1,\ldots,4$, the probability density function (p.d.f.) is expressed as 
$$P(\hat{d}_{A_iT};\boldsymbol{p}_{T})=\frac{1}{\sqrt{2\pi} \sigma_{0}}\exp\left(-\frac{(\hat{d}_{A_iT}-d_{A_iT}(\boldsymbol{p}_{T}))^2}{2\sigma_{0}^2}\right),$$
where $i=1,\ldots,4$. This formulation allows us to transform the localization problem into an unconstrained optimization task:
\begin{equation} \label{model:LF_Case_I_3D}
\begin{split}
\max\limits_{\boldsymbol{p}_{T}} & \prod\limits_{i=1}^{4}P(\hat{d}_{A_iT};{\boldsymbol{p}}_{{T}}).
\end{split}
\end{equation}
By applying the logarithm transformation to the likelihood function in (\ref{model:LF_Case_I_3D}), we obtain:
\begin{equation*}
  L_{I}(\boldsymbol{\hat{d}}_{p};{\boldsymbol{p}}_{{T}})=-\sum\limits_{i=1}^{4}\left(\hat{d}_{A_iT}-d_{A_iT}({\boldsymbol{p}}_{{T}})\right)^2.
\end{equation*}
The maximum likelihood estimate (MLE) $\boldsymbol{\hat{p}}_{{T}}$ is then determined by solving:
\begin{equation}
\boldsymbol{\hat{p}}_{T}=\arg\min\limits_{\boldsymbol{p}_{T}} L_{I}(\boldsymbol{\hat{d}}_{p};{\boldsymbol{p}}_{T})
\label{eq:MLE_q_T_3D}
\end{equation}
The GDOP for this scenario has been previously derived in Eq. (\ref{eq:GDOP_p_T_3D}). 
%

For agent localization, the measurement p.d.f. follows a similar formulation:
$$P(\hat{d}_{A_iB_j};\boldsymbol{\beta}_{B})=\frac{1}{\sqrt{2\pi} \sigma_{0}}\exp\left(-\frac{(\hat{d}_{A_iB_j}-d_{A_iB_j}(\boldsymbol{\beta}_{B}))^2}{2\sigma_{0}^2}\right),$$
where $i,j=1,\ldots,4$. This leads to the formulation of an unconstrained optimization problem for agent localization:
\begin{equation} \label{model:ML_Case_II_3D}
\begin{split}
\max\limits_{\boldsymbol{\beta}_{B}} & \prod\limits_{i,j}P(\hat{d}_{A_iB_j};{\boldsymbol{\beta}}_{B}).
\end{split}
\end{equation}
The corresponding $\log$-likelihood function of (\ref{model:ML_Case_II_3D}) is given by:
\begin{equation*}
  L_{II}(\boldsymbol{\hat{d}}_{B};\boldsymbol{\beta}_{B})=-\sum\limits_{i,j\in\{1,2,3\}}\left(\hat{d}_{A_iB_j}-d_{A_iB_j}(\boldsymbol{\beta}_{B})\right)^2.
\end{equation*}

The MLE $\boldsymbol{\hat{\beta}}_{B}$ is obtained by solving:
\begin{equation}
\boldsymbol{\hat{\beta}}_{B}=\arg\min\limits_{\boldsymbol{\beta}_{B}} L_{II}(\boldsymbol{\hat{d}}_{B};{\boldsymbol{\beta}}_{B})
\label{eq:MLE_beta_B_3D}
\end{equation}

The position and pose GDOP of the target agent-$B$ are derived previously in Eq. (\ref{eq:3D_GDOP_position}), Eq. (\ref{eq:3D_GDOP_gamma}), Eq. (\ref{eq:3D_GDOP_theta}), and Eq. (\ref{eq:3D_GDOP_psi}).
The CRLB and the theoretical best RMSE can be derived from the GDOP. 
For sensor localization, the best achievable RMSE of $\boldsymbol{\hat{p}}_T$ is:
\begin{equation}
\mathrm{CRLB}_{{\boldsymbol{p}}_T}\triangleq\sigma_0\cdot\mathrm{GDOP}_{\boldsymbol{p}_T}.
\label{eq:CRLB_p_T_3D}
\end{equation}
For agent localization, the best achievable RMSE of $\boldsymbol{\hat{p}}_B$ and $\hat{\theta}_B$ are respectively:
\begin{equation}
\mathrm{CRLB}_{{\boldsymbol{p}}_B}\triangleq\sigma_0\cdot\mathrm{GDOP}_{\boldsymbol{p}_B},
  \label{eq:CRLB_q_B_3D}
\end{equation}
\begin{equation}
    \begin{aligned}
    \mathrm{CRLB}_{\gamma_B}&\triangleq\sigma_0\cdot\mathrm{GDOP}_{\gamma_B},\\
    \mathrm{CRLB}_{\theta_B}&\triangleq\sigma_0\cdot\mathrm{GDOP}_{\theta_B},\\
    \mathrm{CRLB}_{\psi_B}&\triangleq\sigma_0\cdot\mathrm{GDOP}_{\psi_B}.
    \end{aligned}    
    \label{eq:CRLB_rpy_B_3D}
\end{equation}
\subsection{Computational Complexity}
\label{subsect:ComputationalComplexity_3D}
In our proposed framework, infrastructure-free relative localization relies solely on onboard sensors and micro-controller units (MCUs). Given the constrained computational resources, a thorough analysis of computational complexity is essential. Let $N$ denote the number of sensors on agent-$A$, and $M$ represent the number of sensors on the target, where $M=1$ for sensor localization and $M=N$ for agent localization. Based on these parameters, we derive the computational complexities for different algorithms in both cases.

\begin{itemize}
    \item \textbf{TT}: The complexity for sensor localization is $\mathcal{O}(N)$, primarily due to matrix multiplication. For agent localization, the complexity increases to $\mathcal{O}(N^2)$ because of the repetitive localization process for the sensors on the target agent.
    
    \item \textbf{Fro-CVX}: The initialization process exhibits a complexity of $\mathcal{O}(N)$ for sensor localization and $\mathcal{O}(N^2)$ for agent localization. During iteration, the complexity remains $\mathcal{O}(N)$ in both cases when employing Newton's method.
    
    \item \textbf{EDMT}: The complexity of EDMT for sensor localization is $\mathcal{O}(N^3)$, driven by the matrix decomposition in Algorithm \ref{alg:EDM_basedPoint_3D}. For agent localization, the complexity is $\mathcal{O}(N^4)$ for EDMT-individually and $\mathcal{O}(N^3)$ for EDMT-jointly.
    
    \item \textbf{MLE}: The initialization complexity depends on whether TT or EDMT is utilized. During iteration, the complexity is $\mathcal{O}(N)$ in both cases when using Newton's method.
\end{itemize}

Given that all methods exhibit polynomial computational complexities and $N$ is limited to $4$ in our framework, the practical difference in computational complexity between algorithms is insignificant. To further validate this, we measured the actual computational time by conducting $10^5$ trials for both sensor and agent localization, with the results documented in Table \ref{tab:Case_I_TimeCost_3D} and Table \ref{tab:Case_II_TimeCost_3D}.

\section{Simulation}
\label{sect:3D_Simulation}
The preceding section primarily addressed performance metrics and localization algorithms. Here, we conduct a simulation-based evaluation of the aforementioned algorithms for near-field relative localization. We model the noise as i.i.d. white Gaussian with a standard deviation of $\SI{5}{\cm}$ in the short-range scenario.

The simulation results demonstrate that the proposed EDMT achieves high accuracy while maintaining low computational complexity for both sensor and agent localization. However, the proposed MLE outperforms other methods, delivering superior performance in both position and heading estimation for agent localization at the cost of higher computational complexity.


\subsection{Algorithm Comparison for Sensor Localization}
\label{subsect:3D_Simulation_Case_I}
In this section, we analyze the relative localization performance of various algorithms for sensor localization. The analysis assumes a sensor configuration tetrahedron with side length $a=\SI{100}{\cm}$, which serves as the foundation for agent localization. The best achievable RMSE, $\mathrm{CRLB}_{\boldsymbol{p}_T}$ in Eq. (\ref{eq:CRLB_p_T_3D}), is adopted as the primary performance metric. The RMSEs of $\boldsymbol{\hat{p}}_T$ obtained using state-of-the-art methods, TT and Fro-CVX, are utilized as benchmarks for comparison.

\begin{figure}[!htbp]
    \centering
    \includegraphics[width=0.4\textwidth]{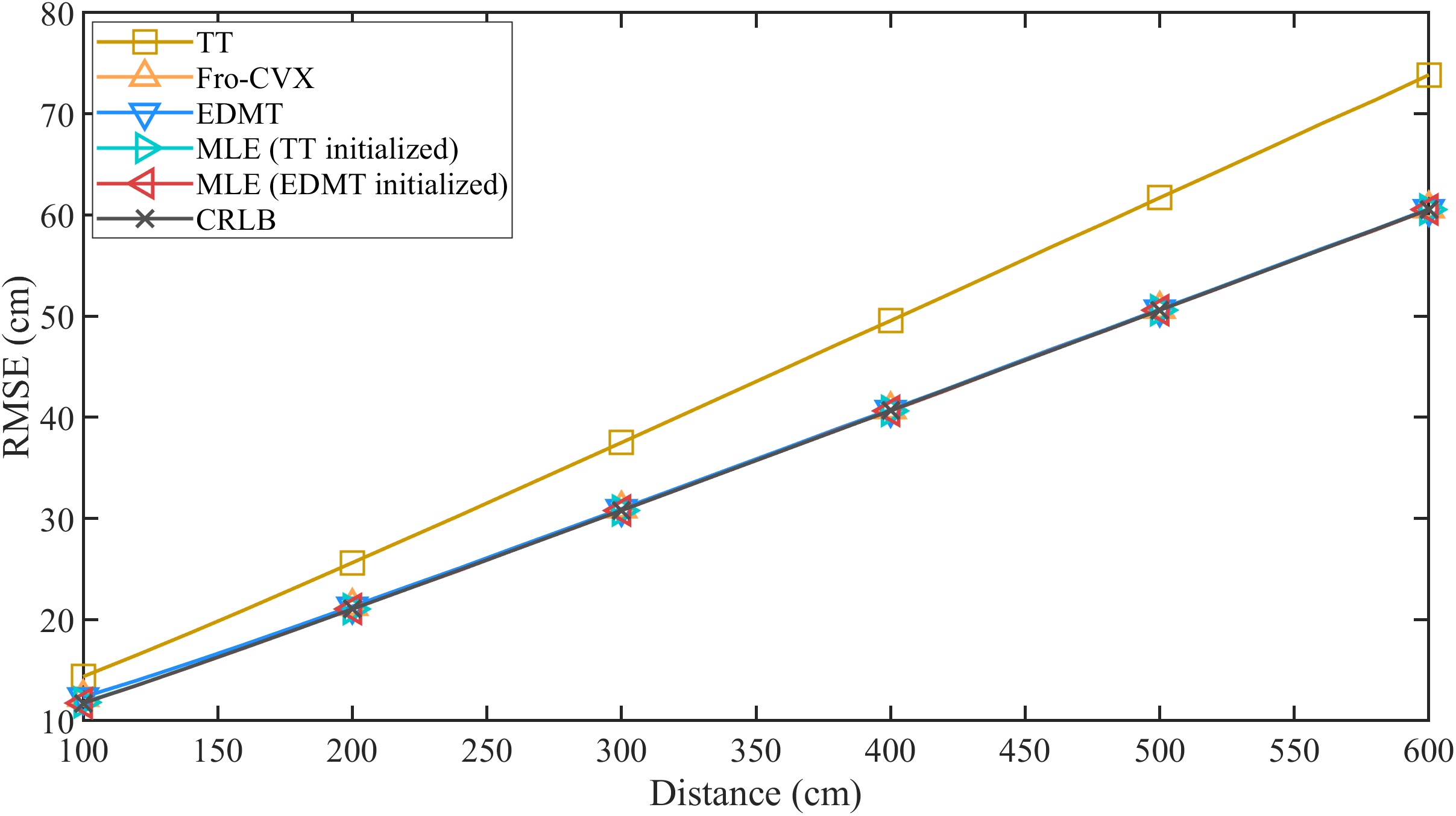}
    \caption{Comparison among the RMSEs of the target sensor position estimate $\boldsymbol{\hat{p}}_{T}$ using EDMT, MLE, TT, and Fro-CVX with the $\mathrm{CRLB}_{\boldsymbol{p}_T}$ when the sensor configuration triangle has its side $a$ equals to $\SI{100}{\cm}$. }
    \label{fig:Comparison_Case_I_RMSExdistance_3D}
\end{figure}

We evaluate the RMSEs of $\boldsymbol{\hat{p}}_T$ for EDMT and MLE against the performance metric $\mathrm{CRLB}_{\boldsymbol{p}_T}$ and the benchmarks across a range of distances from $\SI{100}{\cm}$ to $\SI{600}{\cm}$, as illustrated in Fig.~\ref{fig:Comparison_Case_I_RMSExdistance_3D}. The MLE results are derived from Newton's method initialized with TT and EDMT, respectively. The computational complexities of all methods are compared by measuring the time cost of $10^5$ trials, as detailed in Table~\ref{tab:Case_I_TimeCost_3D}. The proposed EDMT and MLE achieve RMSEs that surpass the benchmark TT and closely approach $\mathrm{CRLB}_{\boldsymbol{p}_T}$, while maintaining computational complexity comparable to Fro-CVX. Based on the computational time cost analysis in Table~\ref{tab:Case_I_TimeCost_3D}, it is evident that EDMT and Fro-CVX represent the optimal choices for sensor localization within the proposed framework.



\begin{figure*}[!htbp]
     \centering
     \begin{subfigure}[]{0.32\textwidth}
         \centering
         \includegraphics[width=\textwidth]{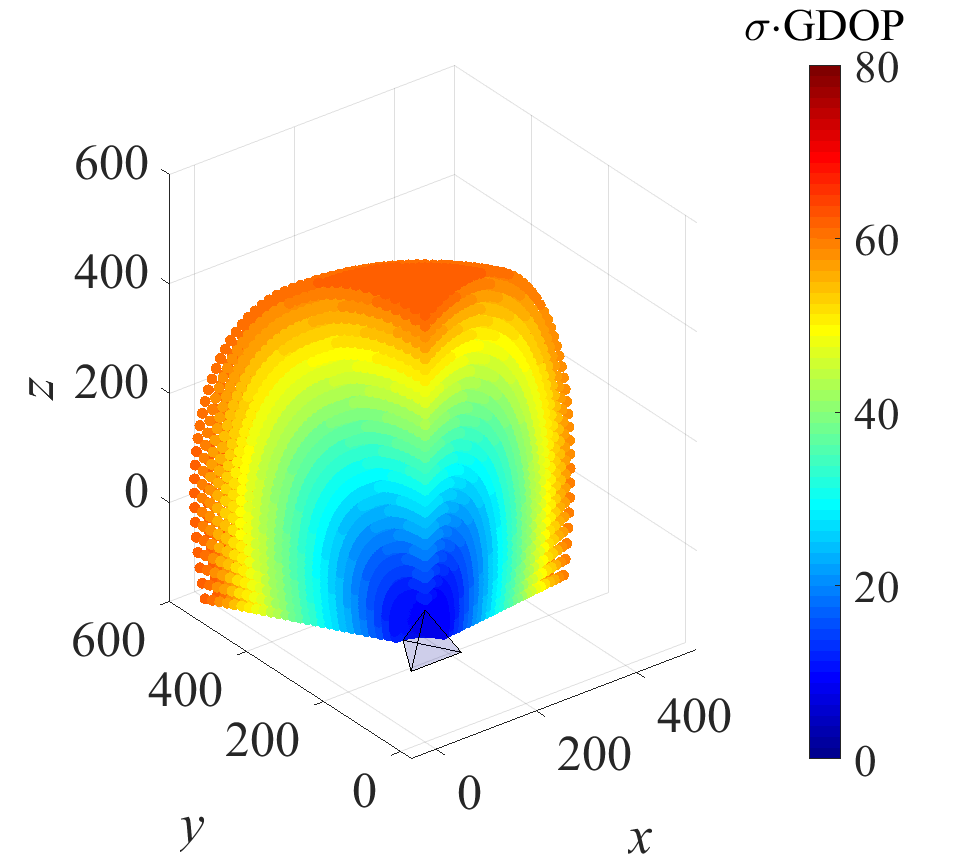}
         \caption{CRLB}
         \label{fig:CRLB_A100_3D}
     \end{subfigure}
     \begin{subfigure}[]{0.32\textwidth}
         \centering
         \includegraphics[width=\textwidth]{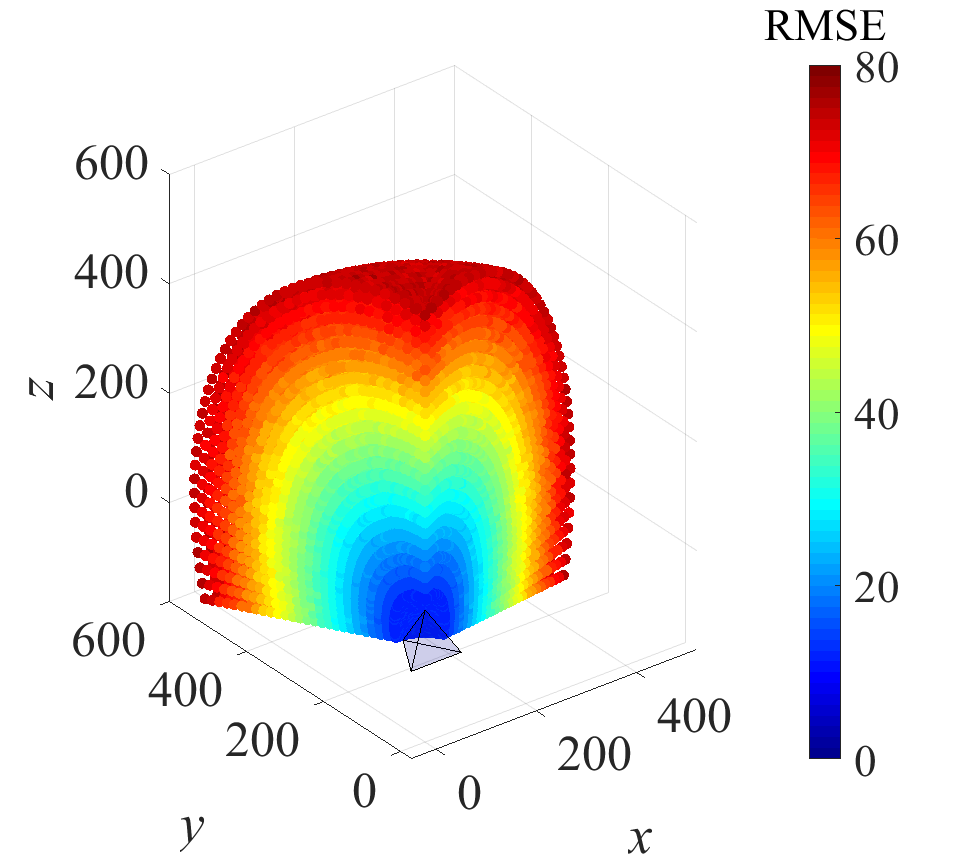}
         \caption{TT}
         \label{fig:TT_A100_3D}
     \end{subfigure}
     \begin{subfigure}[]{0.32\textwidth}
         \centering
         \includegraphics[width=\textwidth]{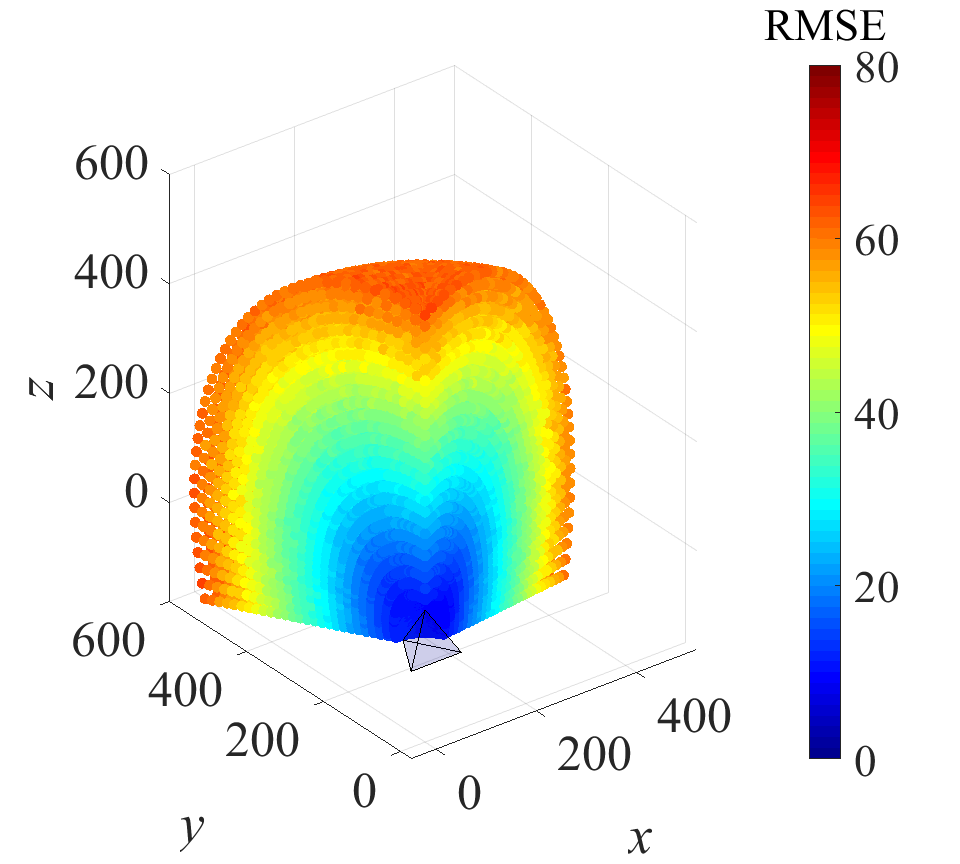}
         \caption{Fro-CVX}
         \label{fig:Fro_CVX_A100_3D}
     \end{subfigure}\\ 
      \begin{subfigure}[]{0.32\textwidth}
         \centering
         \includegraphics[width=\textwidth]{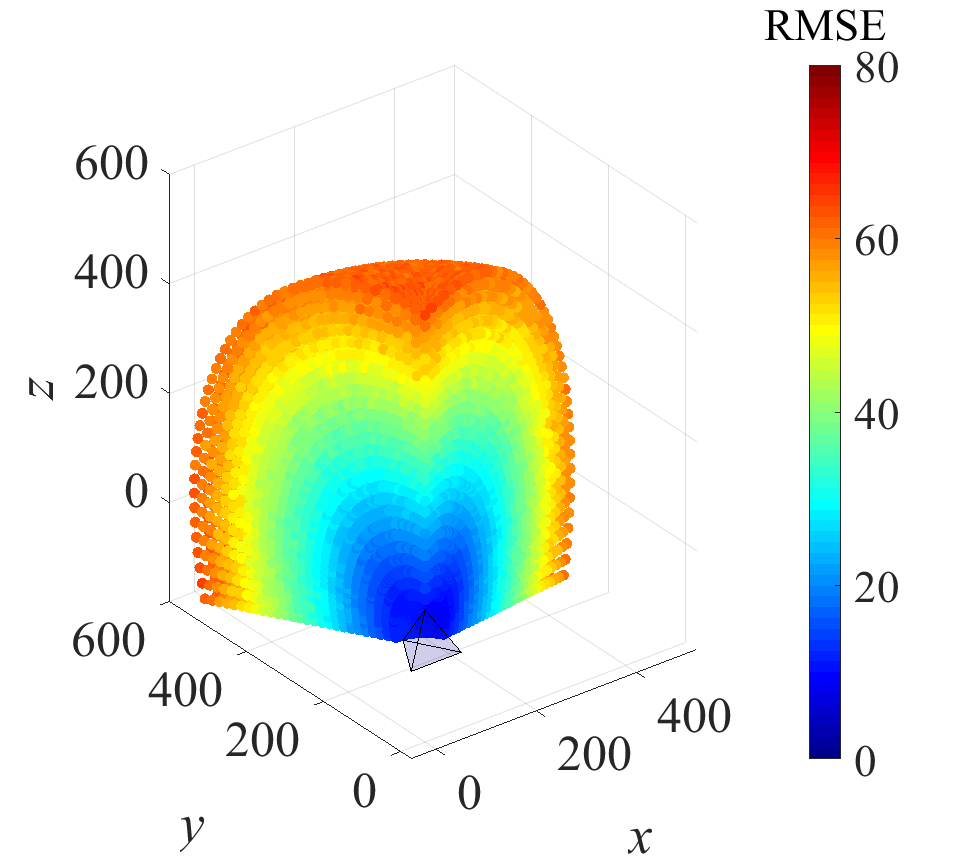}
         \caption{EDMT}
         \label{fig:EDMT_A100_3D}
     \end{subfigure}
      \begin{subfigure}[]{0.32\textwidth}
         \centering
         \includegraphics[width=\textwidth]{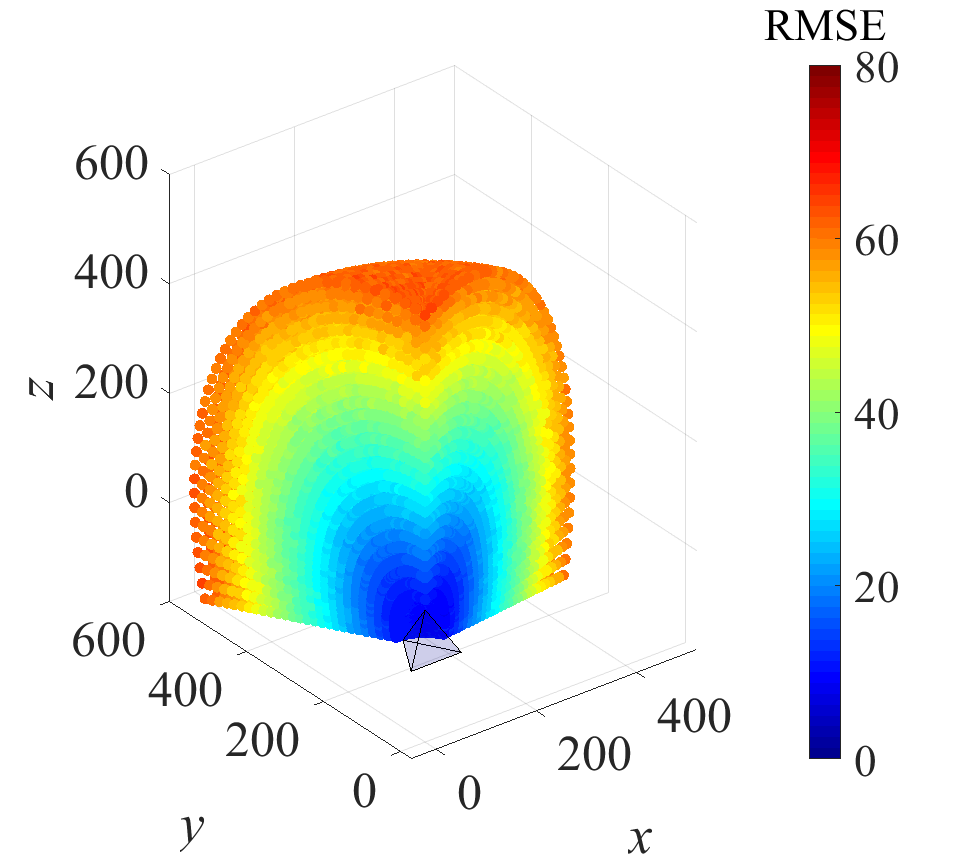}
         \caption{MLE (TT initialized)}
         \label{fig:MLE_TT_A100_3D}
     \end{subfigure} 
     \begin{subfigure}[]{0.32\textwidth}
         \centering
         \includegraphics[width=\textwidth]{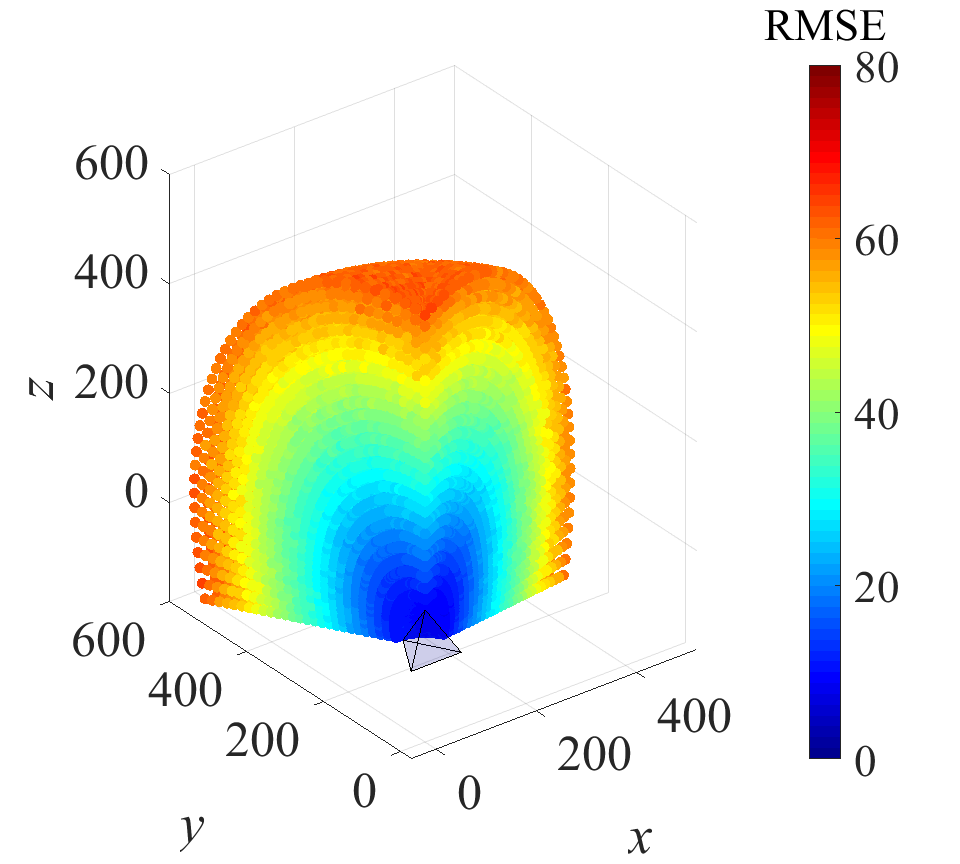}
         \caption{MLE (EDMT initialized)}
         \label{fig:MLE_EDMT_A100_3D}
     \end{subfigure}
  \caption{Average RMSEs of a target sensor in $1000$ simulation runs when the range measurement noise is i.i.d. Gaussian with a standard deviation $\sigma=\SI{5}{\cm}$, and the configuration triangle has side $a=\SI{100}{\cm}$. }
  \label{fig:Case_I_RMSE_Distribution_3D}
\end{figure*}

To provide a more comprehensive analysis, we compare the detailed RMSE distributions of EDMT and MLE with those of TT, Fro-CVX, and the theoretical lower bound $\mathrm{CRLB}_{\boldsymbol{p}_T}$. This comparison is conducted within a spherical cone space defined by a radial distance ranging from \SI{80}{\cm} to \SI{500}{\cm}, a polar angle spanning \SI{0}{\degree} to \SI{180}{\degree}, and an azimuthal angle between \SI{30}{\degree} and \SI{90}{\degree} in a spherical coordinate system, as depicted in Fig.~\ref{fig:Case_I_RMSE_Distribution_3D}. The 3D contour graph of the RMSE distribution reveals that both the proposed EDMT and MLE exhibit robust estimation performance, with their RMSE distributions closely approaching the theoretical lower bound. Moreover, it is crucial to emphasize that when the RMSE for sensor localization approaches half the sensor configuration triangle side length $a$, the position can still be estimated, but the yaw, pitch, and roll angles for agent localization cannot be determined. This limitation arises because the sensors on the target agent become indistinguishable under such conditions.


\begin{table}[!htbp]
 \caption{The time cost comparison of different algorithms in 3D Case I taking $10^5$ trials.}
 \label{tab:Case_I_TimeCost_3D}
 \centering
 \begin{tabular}{c c c c}
 \toprule[1.25pt]
 \textbf{Algorithm} & \textbf{Time (s)} & \textbf{Ratio} & \textbf{Complexity}\\
 \midrule 
 TT & 1.07 & \textbf{1} & $\mathcal{O}(N)$\\
 Fro-CVX & 3.46 & 3.22 & $\mathcal{O}(N)$\\
 EDMT & 4.11 & 3.83 & $\mathcal{O}(N^3)$\\
  MLE (TT initialized) & 11.32 & 10.53 & $\mathcal{O}(N)$\\
  MLE (EDMT initialized) & 14.17 & 13.18 & $\mathcal{O}(N^3)$\\
 \bottomrule [1.25pt]
 \end{tabular}
\end{table}

\subsection{Algorithm Comparison in Case II}
\label{subsect:3D_Simulation_Case_II}
In this section, we evaluate the performance of various relative localization algorithms for agent localization when the sensor configuration tetrahedron has a side length of \( a = \SI{100}{\cm} \), and the target agent is positioned at a range varying from \(\SI{200}{\cm}\) to \(\SI{500}{\cm}\). The theoretical lower bounds, \(\mathrm{CRLB}_{\boldsymbol{p}_B}\), \(\mathrm{CRLB}_{\gamma_B}\), \(\mathrm{CRLB}_{\theta_B}\), and \(\mathrm{CRLB}_{\psi_B}\), as defined in Eq. (\ref{eq:CRLB_q_B_3D}) and Eq. (\ref{eq:CRLB_rpy_B_3D}), are employed as performance metrics. Additionally, the RMSEs of the position estimate \(\boldsymbol{\hat{p}}_B\) and orientation estimates \(\hat{\gamma}_{B}\), \(\hat{\theta}_{B}\), and \(\hat{\psi}_{B}\), obtained using state-of-the-art methods (TT and Fro-CVX), serve as benchmarks for comparison.

\begin{figure}[!htbp]
    \centering
    \includegraphics[width=0.4\textwidth]{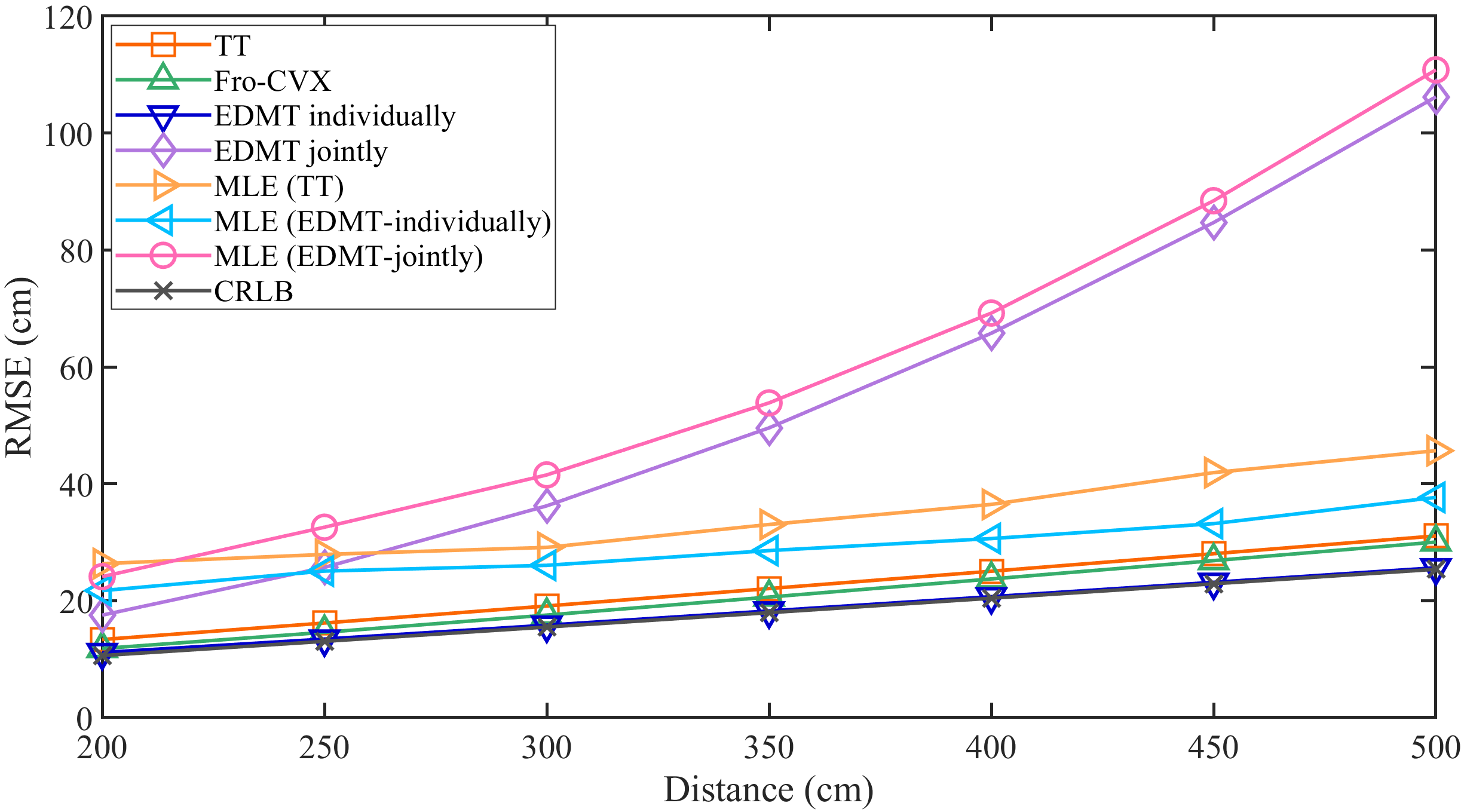}
    \caption{Comparison among the RMSEs of the target sensor position estimate $\boldsymbol{\hat{p}}_{B}$ using EDMT, MLE, TT, and Fro-CVX with the $\mathrm{CRLB}_{\boldsymbol{p}_B}$ when the sensor configuration triangle has its side $a$ equals to $\SI{100}{\cm}$. }
    \label{fig:Comparison_Case_II_RMSExdistance_3D}
\end{figure}

We specifically compare the RMSEs of \(\boldsymbol{\hat{p}}_B\) achieved by the proposed methods, EDMT and MLE, against \(\mathrm{CRLB}_{\boldsymbol{p}_B}\) and the benchmark methods, as illustrated in Fig.~\ref{fig:Comparison_Case_II_RMSExdistance_3D}. The results show that the RMSE of the position estimate using the proposed EDMT-individually method nearly attains the theoretical lower bound, demonstrating superior performance compared to the other methods. 


\begin{figure}[!htbp]
    \centering
    \includegraphics[width=0.4\textwidth]{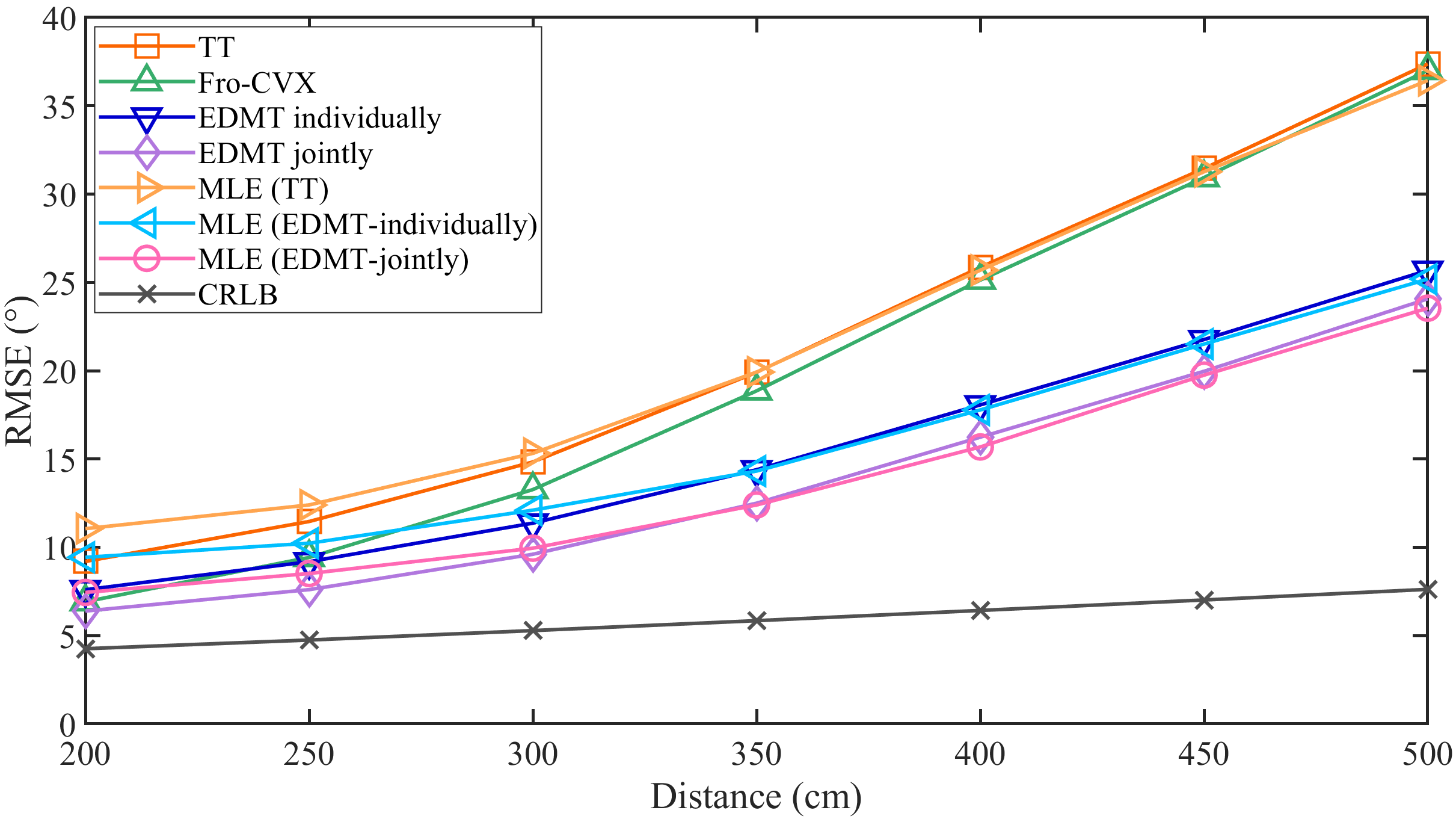}
    \caption{Comparison among the RMSEs of the target sensor position estimate $\hat{\gamma}_B$ using EDMT, MLE, TT, and Fro-CVX with the $\mathrm{CRLB}_{\gamma_B}$ when the sensor configuration triangle has its side $a$ equals to $\SI{100}{\cm}$. }
    \label{fig:Compa_Case_II_RMSE_gammaBxdistance_3D}
\end{figure}

\begin{figure}[!htbp]
    \centering
    \includegraphics[width=0.4\textwidth]{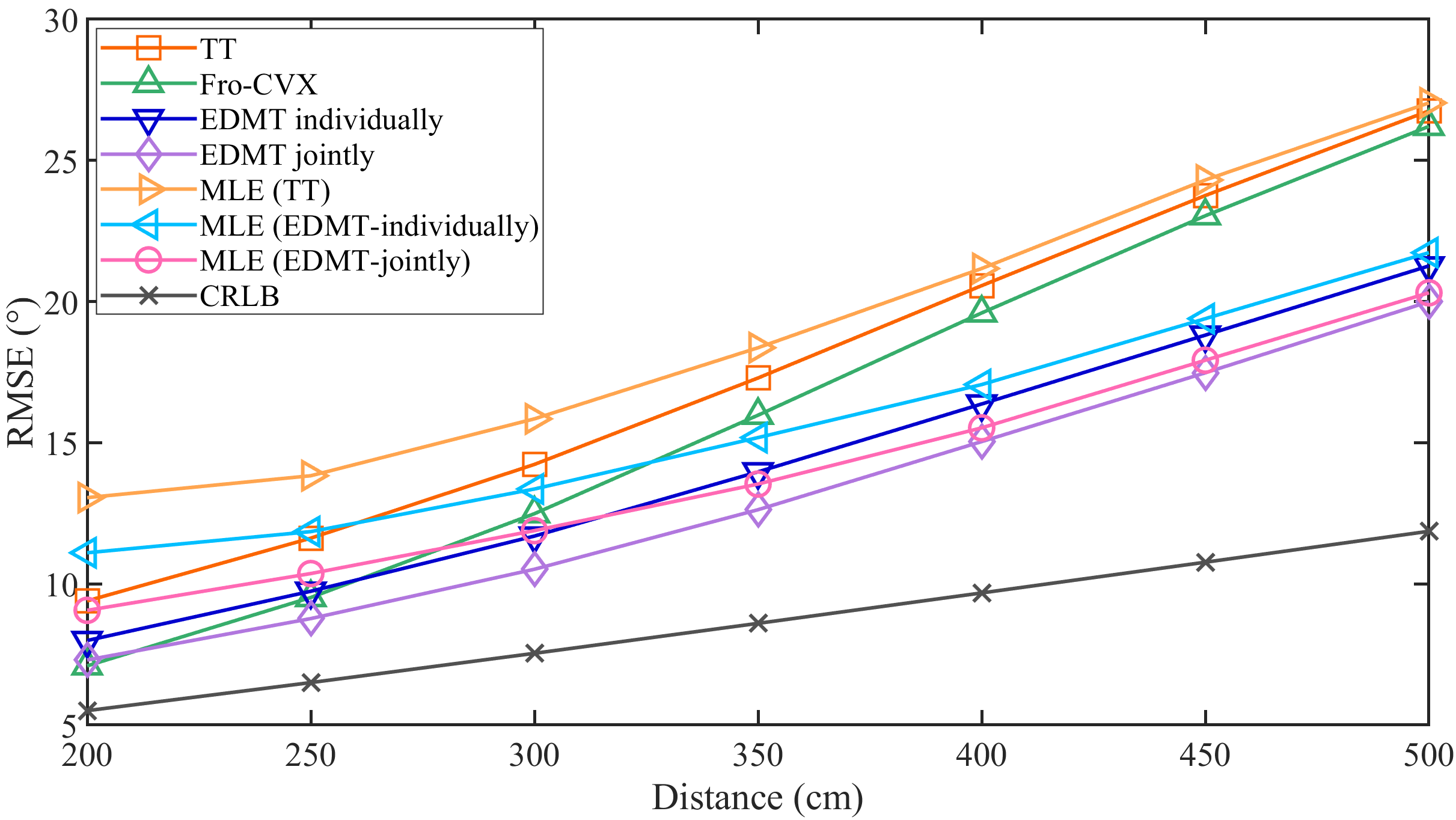}
    \caption{Comparison among the RMSEs of the target sensor position estimate $\hat{\theta}_B$ using EDMT, MLE, TT, and Fro-CVX with the $\mathrm{CRLB}_{\theta_B}$ when the sensor configuration triangle has its side $a$ equals to $\SI{100}{\cm}$. }
    \label{fig:Compa_Case_II_RMSE_thetaBxdistance_3D}
\end{figure}

\begin{figure}[!htbp]
    \centering
    \includegraphics[width=0.4\textwidth]{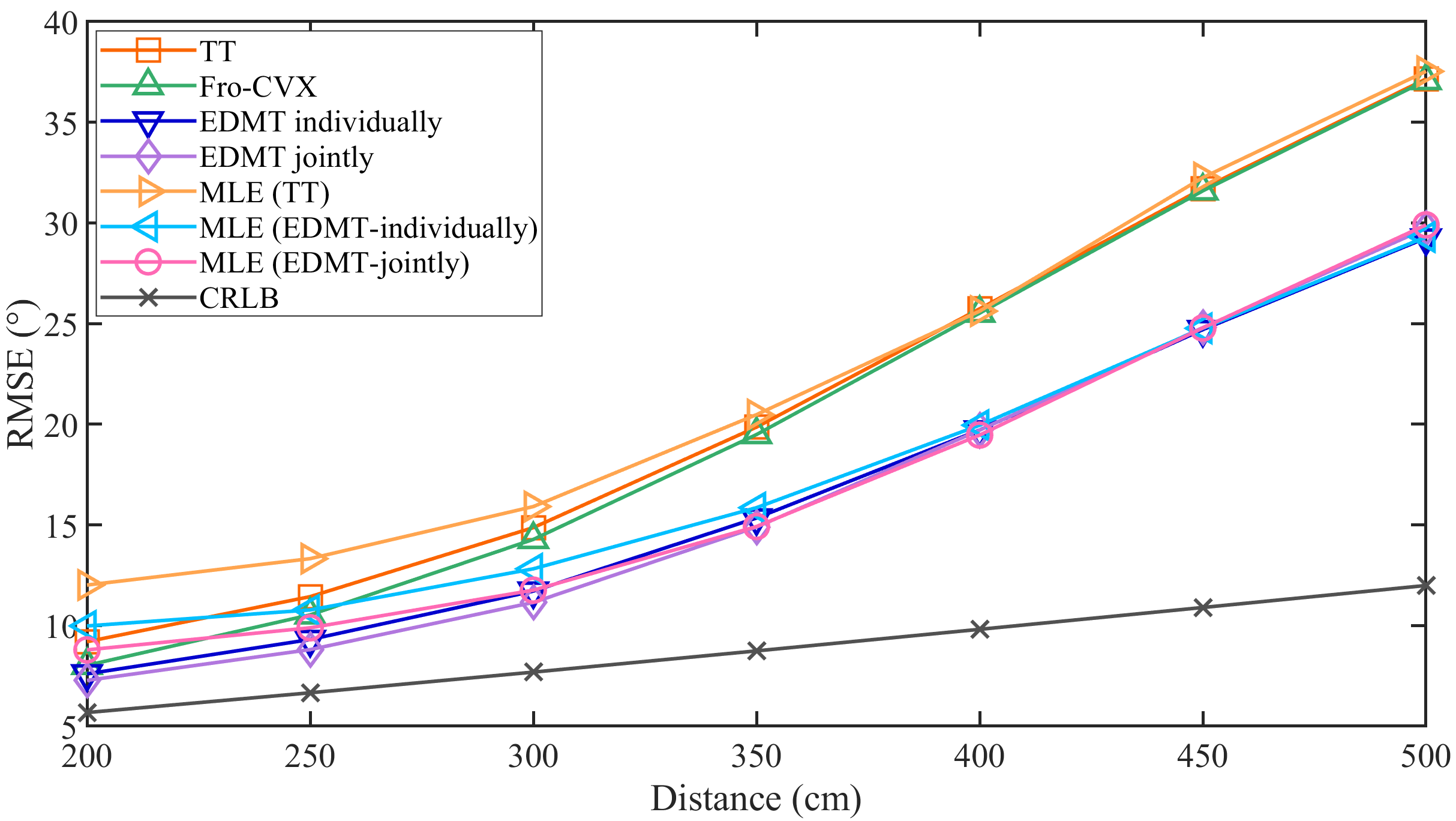}
    \caption{Comparison among the RMSEs of the target sensor position estimate $\hat{\psi}_B$ using EDMT, MLE, TT, and Fro-CVX with the $\mathrm{CRLB}_{\psi_B}$ when the sensor configuration triangle has its side $a$ equals to $\SI{100}{\cm}$. }
    \label{fig:Compa_Case_II_RMSE_psiBxdistance_3D}
\end{figure}

When considering the attitude angle estimates, we compare the RMSEs of the roll, pitch, and yaw angles, namely \(\hat{\gamma}_B\), \(\hat{\theta}_B\), and \(\hat{\psi}_B\), obtained using the proposed methods against the corresponding CRLBs and the RMSEs of state-of-the-art methods. These comparisons are illustrated in Fig.~\ref{fig:Compa_Case_II_RMSE_gammaBxdistance_3D}, Fig.~\ref{fig:Compa_Case_II_RMSE_thetaBxdistance_3D}, and Fig.~\ref{fig:Compa_Case_II_RMSE_psiBxdistance_3D}. The results show that the EDMT-jointly method and MLE initialized with EDMT-jointly perform the best, achieving lower RMSEs compared to the state-of-the-art methods and closely approaching the CRLBs. 
To assess the computational efficiency of the methods, the time required for \(10^4\) trials is measured and presented in Table~\ref{tab:Case_II_TimeCost_3D}. 



\begin{table}[!htbp]
  \caption{The time cost comparison of different algorithms in Case II taking $10^4$ trials}
  \label{tab:Case_II_TimeCost_3D}
  \centering
  \begin{tabular}{c c c c}
\toprule [1.25pt]
  \textbf{Algorithm} & \textbf{Time (s)} & \textbf{Ratio} & \textbf{Complexity}\\
  \midrule 
  TT & 0.37 & \textbf{1} & $\mathcal{O}(N^2)$ \\
  Fro-CVX & 622.6 & 1664.8 & $\mathcal{O}(N^2)$ \\
  EDMT-jointly & 0.73 & 1.96 & $\mathcal{O}(N^3)$ \\
  EDMT-individually & 2.16 & 5.78 & $\mathcal{O}(N^4)$ \\
  MLE (TT initialized) & 1276.4 & 3412.8 & $\mathcal{O}(N^2)$\\
    MLE (EDMT-jointly initialized) & 1208.0 & 3229.9 & $\mathcal{O}(N^3)$ \\
  MLE (EDMT-individually initialized) & 1262.5 & 3375.7 & $\mathcal{O}(N^4)$\\
\bottomrule [1.25pt]
  \end{tabular}
\end{table}

In summary, the proposed EDMT-individually algorithm achieves high accuracy and low computational complexity for estimating the position of a target agent. While the Fro-CVX and MLE methods perform well with acceptable computational complexity in 2D scenarios, both exhibit significantly higher computational demands in 3D scenarios. For position estimation, the proposed EDMT-individually method stands out as the best-performing approach, offering minimal computational complexity. Additionally, while additional sensors are required for attitude angle estimation, the RMSEs of these estimates do not closely approach the CRLB. Therefore, using substitute sensors such as IMUs and magnetometers is a more effective choice for 3D agent attitude estimation. 

\section{Experiments and Applications}
\label{sect:3D_Experiment}
Section~\ref{sect:3D_Simulation} conducts a detailed simulation analysis to evaluate the accuracy and efficiency of the proposed localization algorithms. Leveraging these simulation results, the algorithms are further implemented in hardware experiments to validate their performance in real-world applications. This section begins with an introduction to the hardware platform. A static experiment focused on sensor localization is then performed to verify the conclusions derived from the simulations. Finally, the experimental results are compared with the simulation results, and a thorough analysis is presented. 
\subsection{Field Tests}
\begin{figure}[!htbp]
  \centering
  \includegraphics[width=0.35\textwidth]{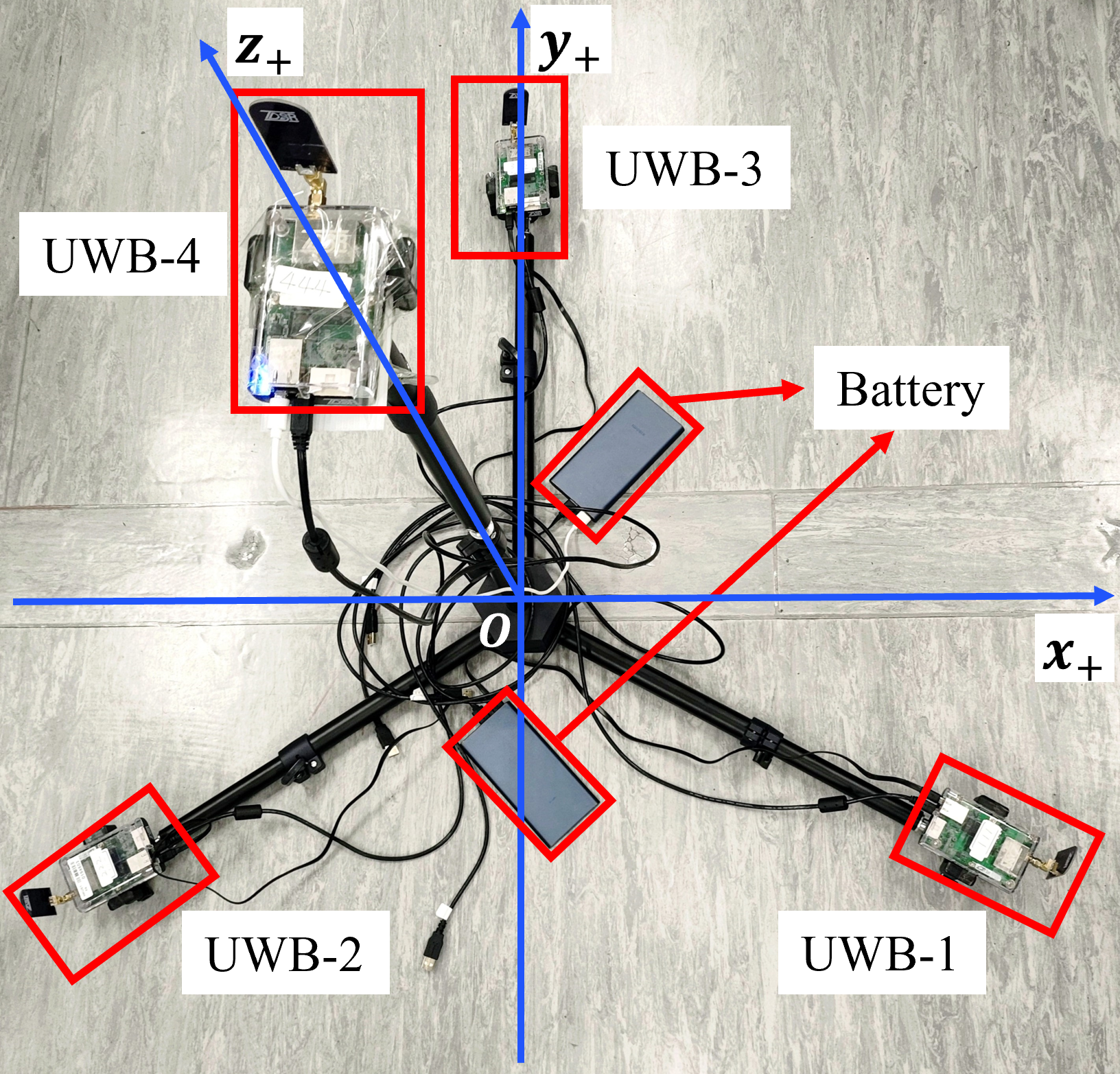}
  \caption{The UWB sensors configured on a 3D agent.}
  \label{fig:HardwarePlatform_3D}
\end{figure}

\begin{figure}[!htbp]
\centering
\includegraphics[width=0.3\textwidth]{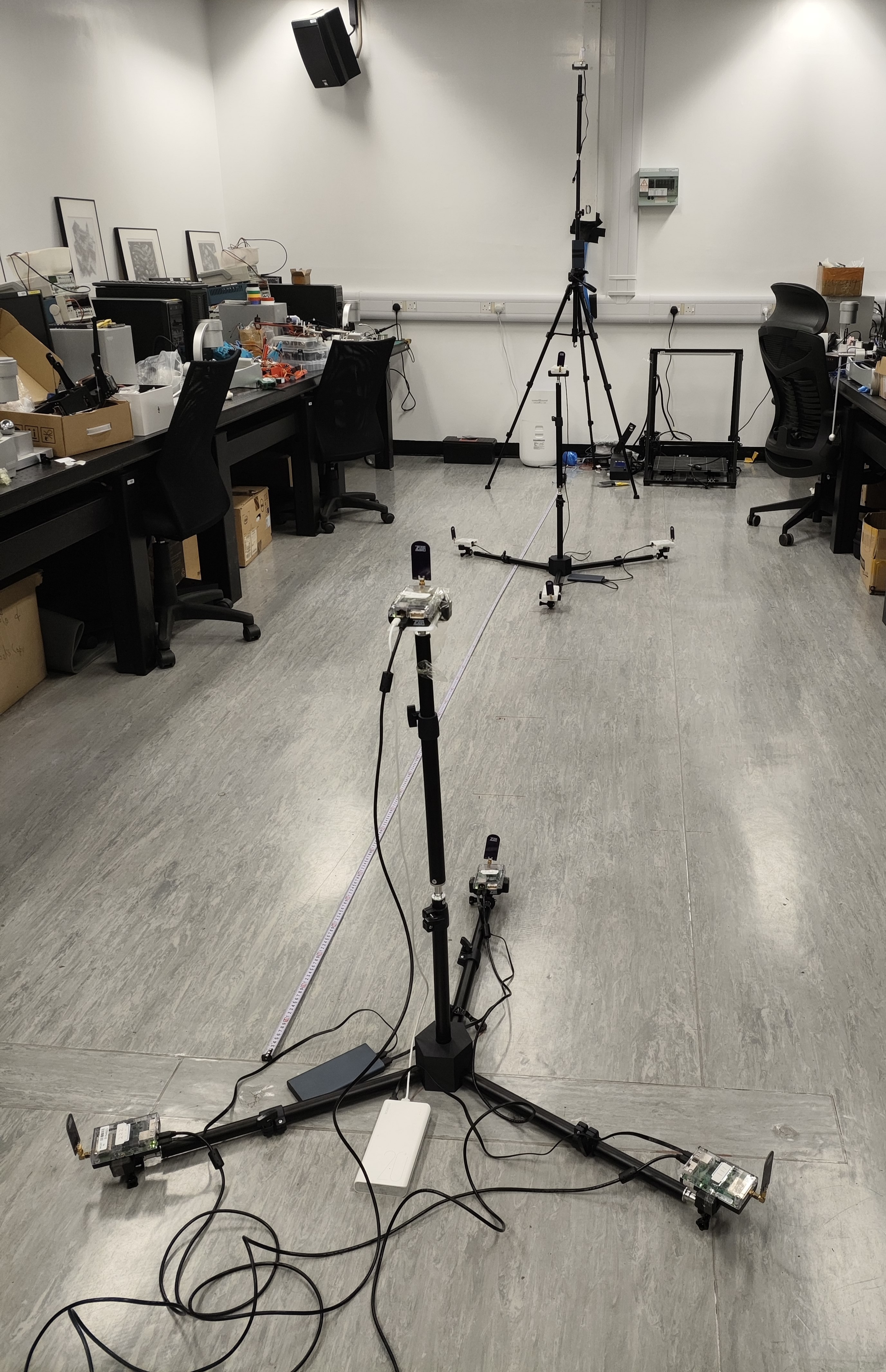}
\caption{The experiment to localize a sensor in 3D scenarios.}
\label{fig:Point_Experiment_3D}
\end{figure}

In our hardware experiment, an agent is constructed using a regular tetrahedron platform equipped with four UWB ranging sensors positioned at its vertices, as illustrated in Fig.~\ref{fig:HardwarePlatform_3D}. Time Domain UWB sensors are employed for this setup. The side length of the regular tetrahedron platform is set to \(\SI{100}{\cm}\), ensuring consistency with the previous configurations. 
Static experiments are conducted for 3D sensor localization at a distance ranging from \(\SI{200}{\cm}\) to \(\SI{500}{\cm}\), as depicted in Fig.~\ref{fig:Point_Experiment_3D}. The relative orientations are configured at \(\SI{0}{\degree}\), \(\SI{30}{\degree}\), and \(\SI{60}{\degree}\) relative to the onboard coordinate system. 

\begin{figure}[!htbp]
        \centering
        \includegraphics[width=0.4\textwidth]{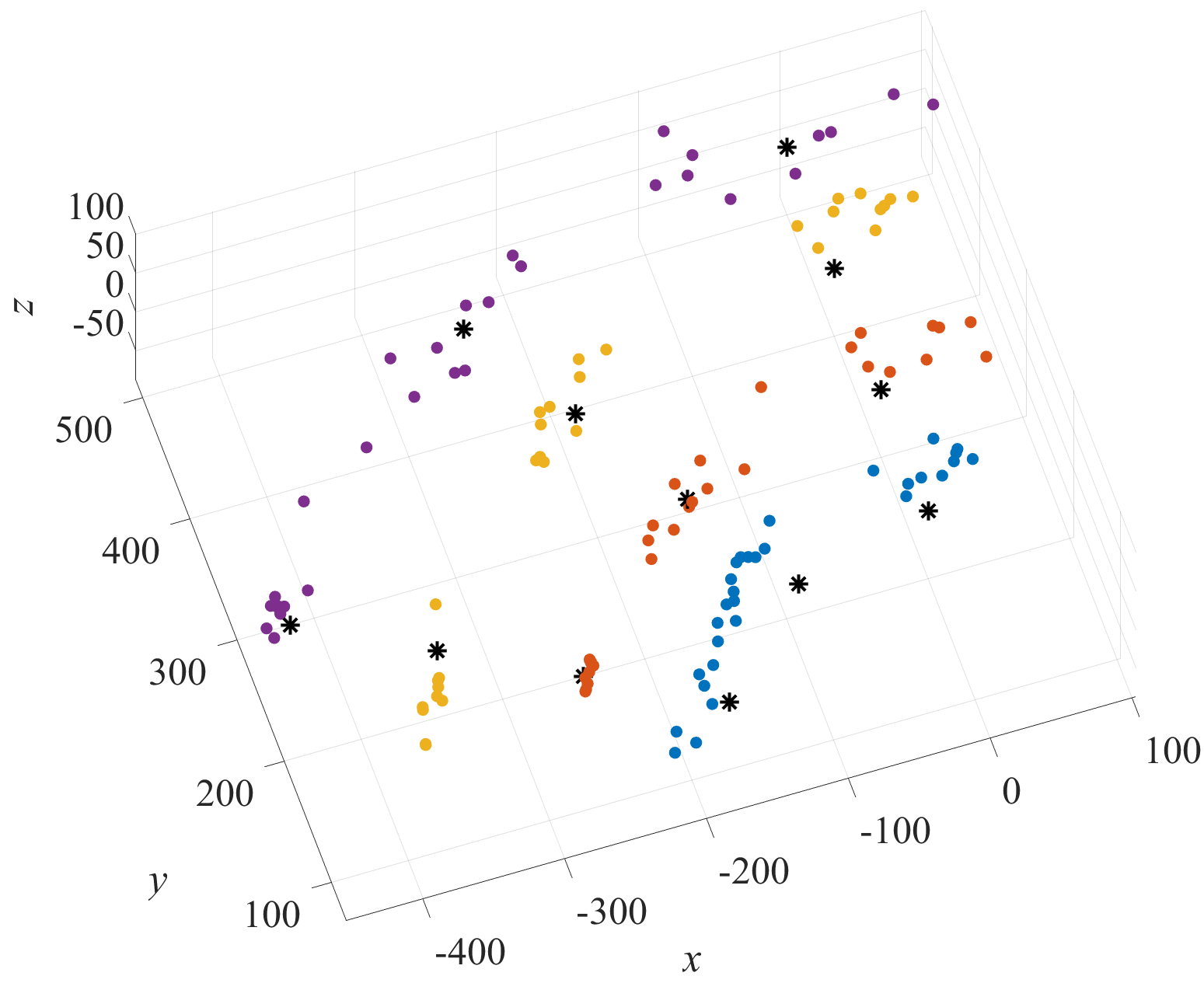}
        \caption{Hardware results to localize a target sensor in 3D scenarios.}
        \label{fig:Hardware_AgentxPoint_3D}
\end{figure}

\begin{figure}[!htbp]
\centering
\includegraphics[width=0.4\textwidth]{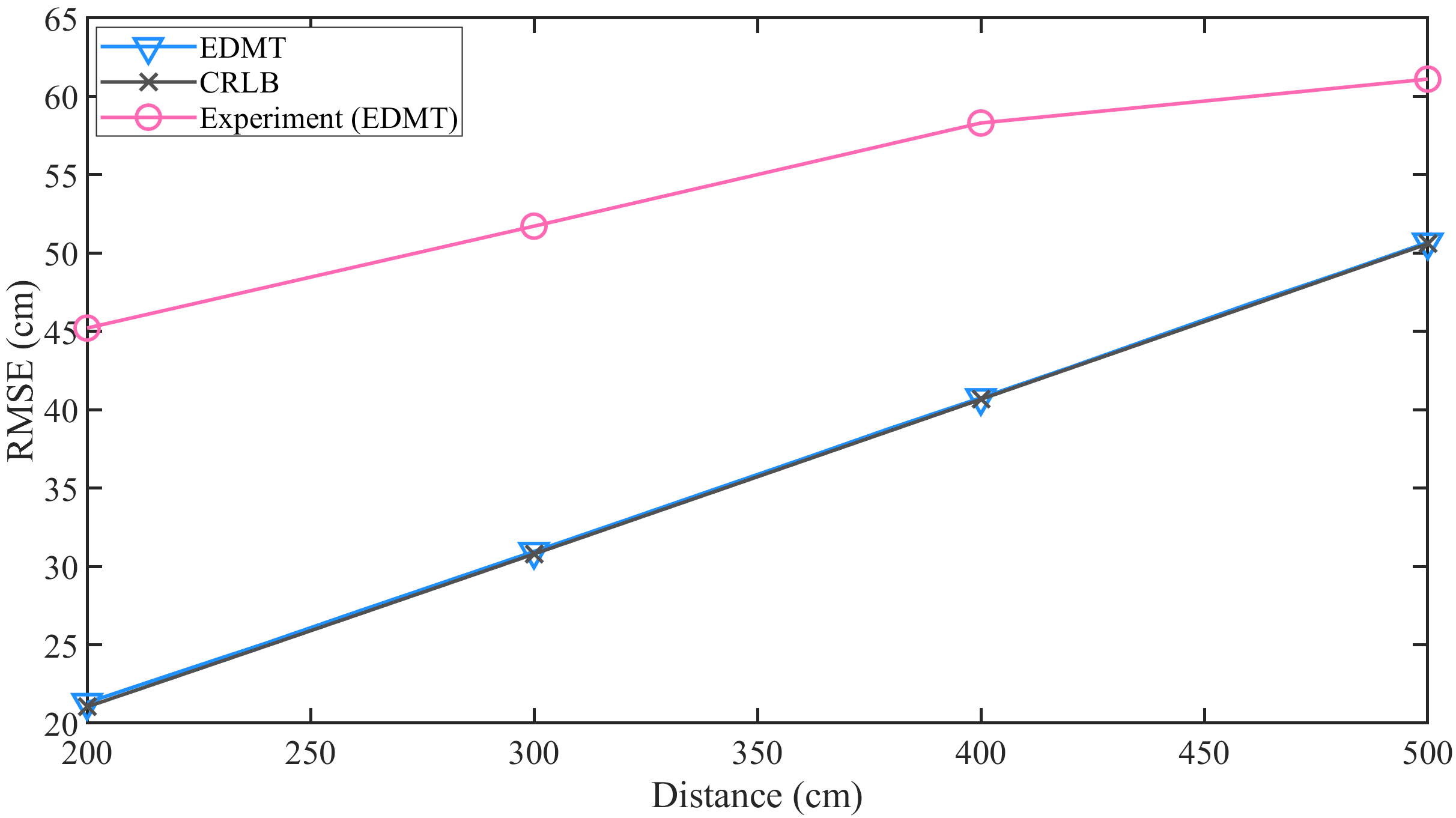}
\caption{The experiment to localize an agent in 3D scenarios.}
\label{fig:Point_Experiment_3D_Results_Compare_CRLB}
\end{figure}

The position estimation results obtained using the proposed EDMT algorithm are compared with the CRLB and the actual positions of the target sensors, as shown in Fig.~\ref{fig:Hardware_AgentxPoint_3D}. Additionally, the RMSE results of the position estimates using the proposed EDMT algorithm are compared with the simulation results, as illustrated in Fig.~\ref{fig:Point_Experiment_3D_Results_Compare_CRLB}. 

\begin{figure}[!htbp]
    \centering
    \begin{subfigure}[b]{0.5\linewidth}
        \centering
        \includegraphics[width=\textwidth]{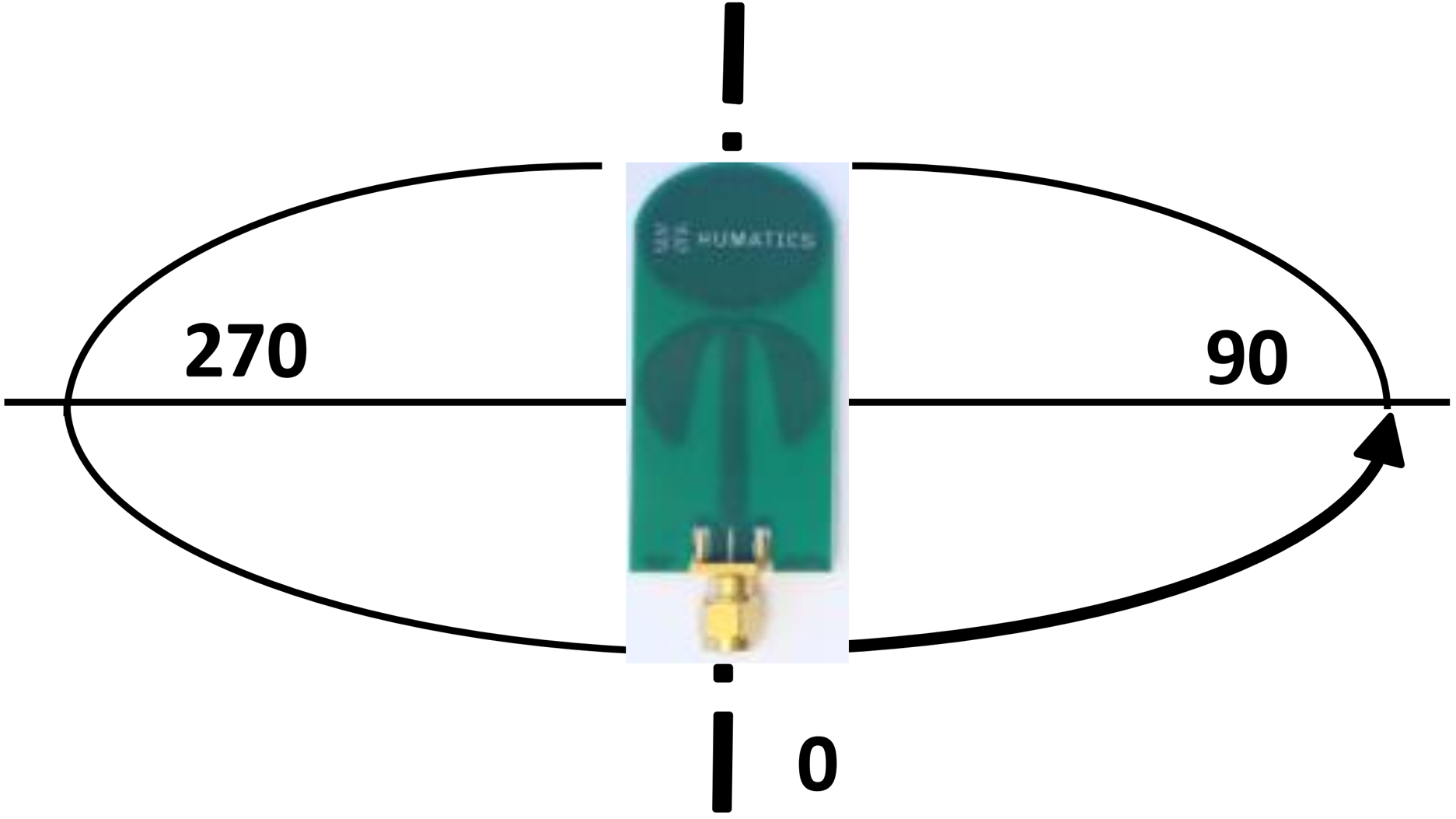}
    \captionsetup{justification=centering}
        \caption{}
        \label{fig:UWB_Antenna_Azimuth_Beam_Pattern_4GHz_Illustration}
    \end{subfigure}\\
    \begin{subfigure}[b]{0.5\linewidth}
        \centering
        \includegraphics[width=\textwidth]{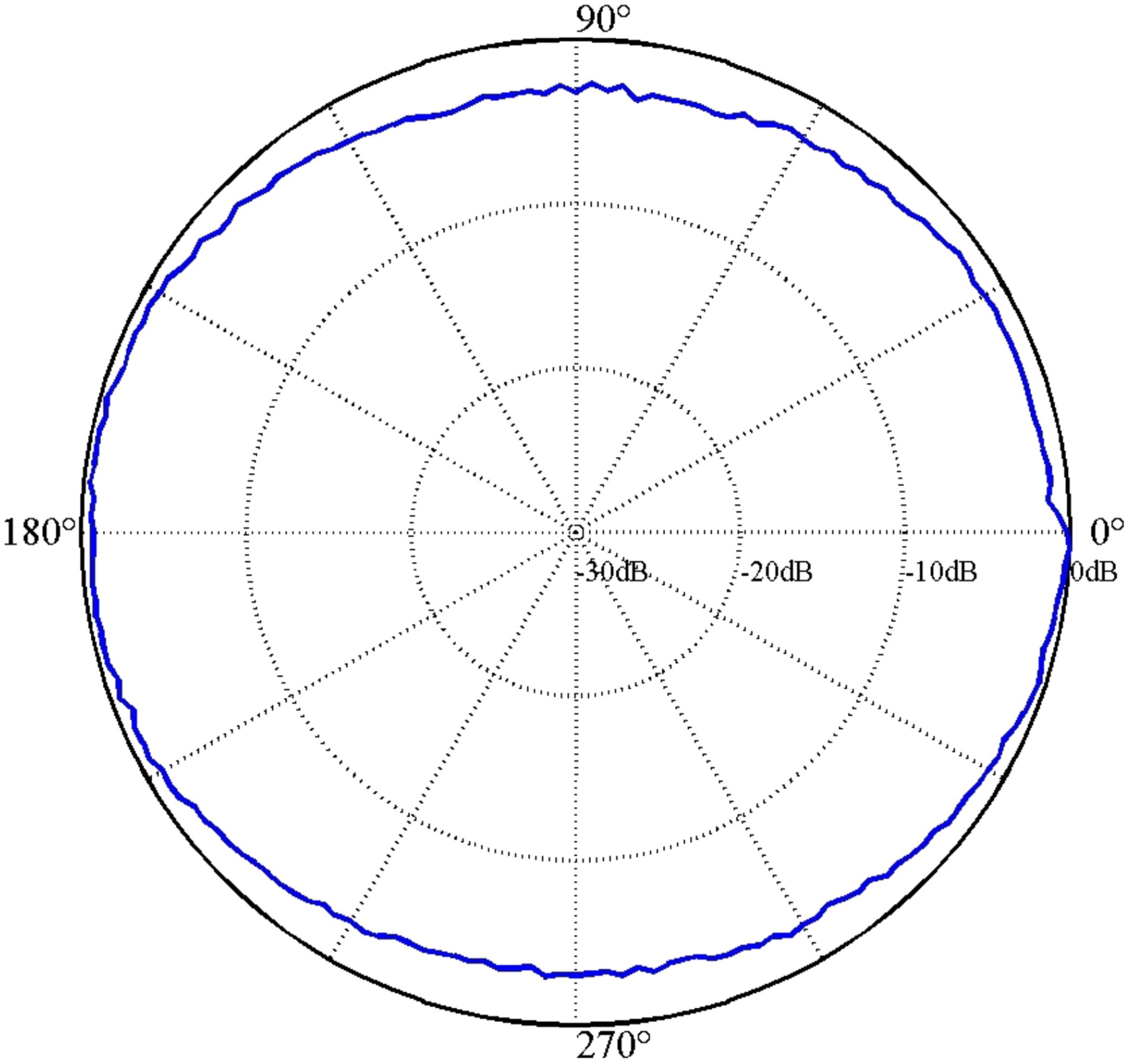}
    \captionsetup{justification=centering}
        \caption{}
        \label{fig:UWB_Antenna_Azimuth_Beam_Pattern_4GHz}
    \end{subfigure}
    \caption{UWB Azimuth Beam Pattern for $\SI{4}{\GHz}$ \cite{timedomain_antenna_manual}.}
    \label{fig:UWB_Antenna_Azimuth_Beam_Pattern}
\end{figure}

\begin{figure}[!htbp]
    \centering
    \begin{subfigure}[b]{0.3\linewidth}
        \centering  
        \includegraphics[width=\textwidth]{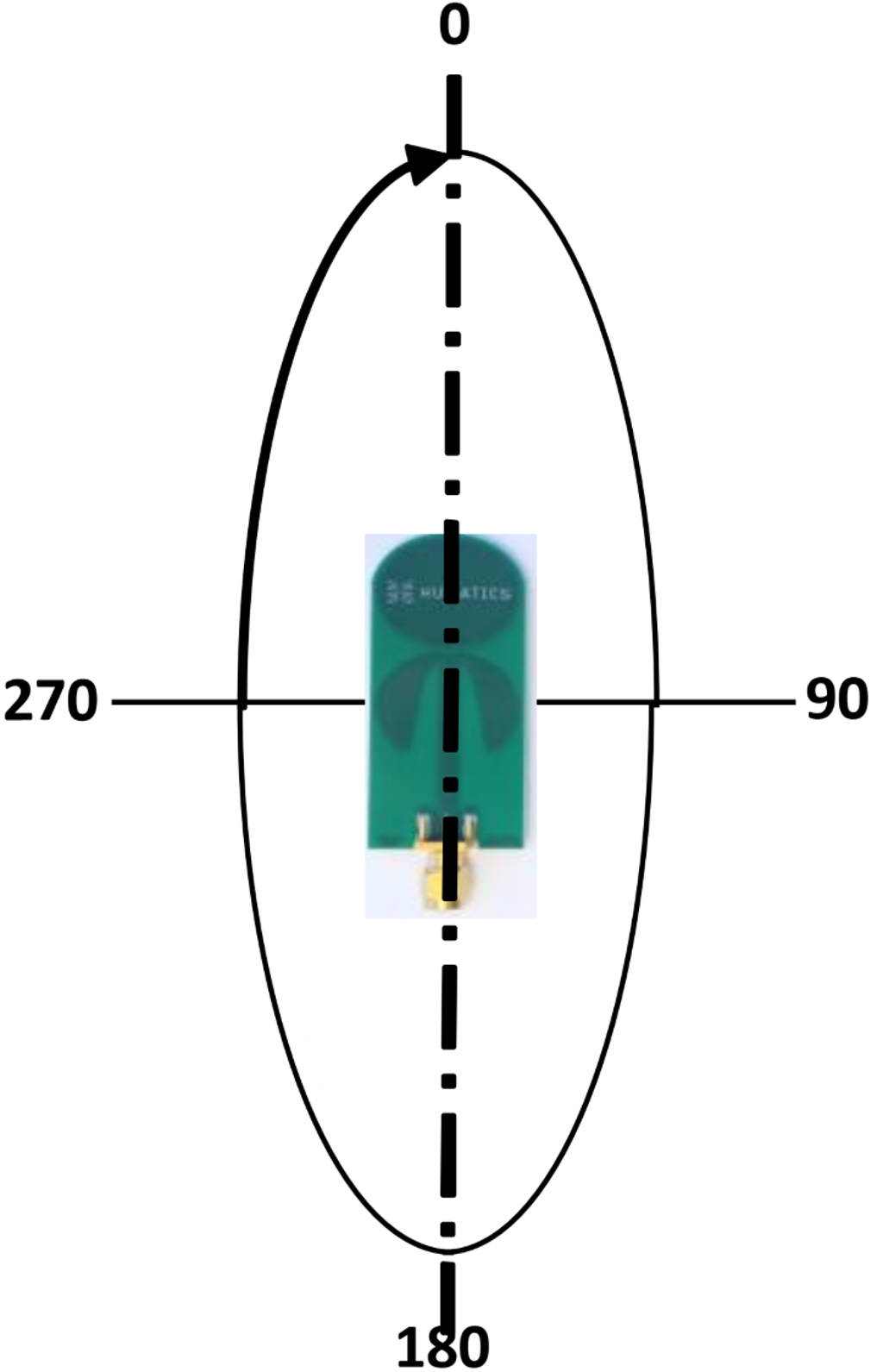}
    \captionsetup{justification=centering}
        \caption{}
        \label{fig:UWB_Antenna_Elevation_Beam_Pattern_4GHz_Illustration}
    \end{subfigure}
    \begin{subfigure}[b]{0.5\linewidth}
        \centering
        \includegraphics[width=\textwidth]{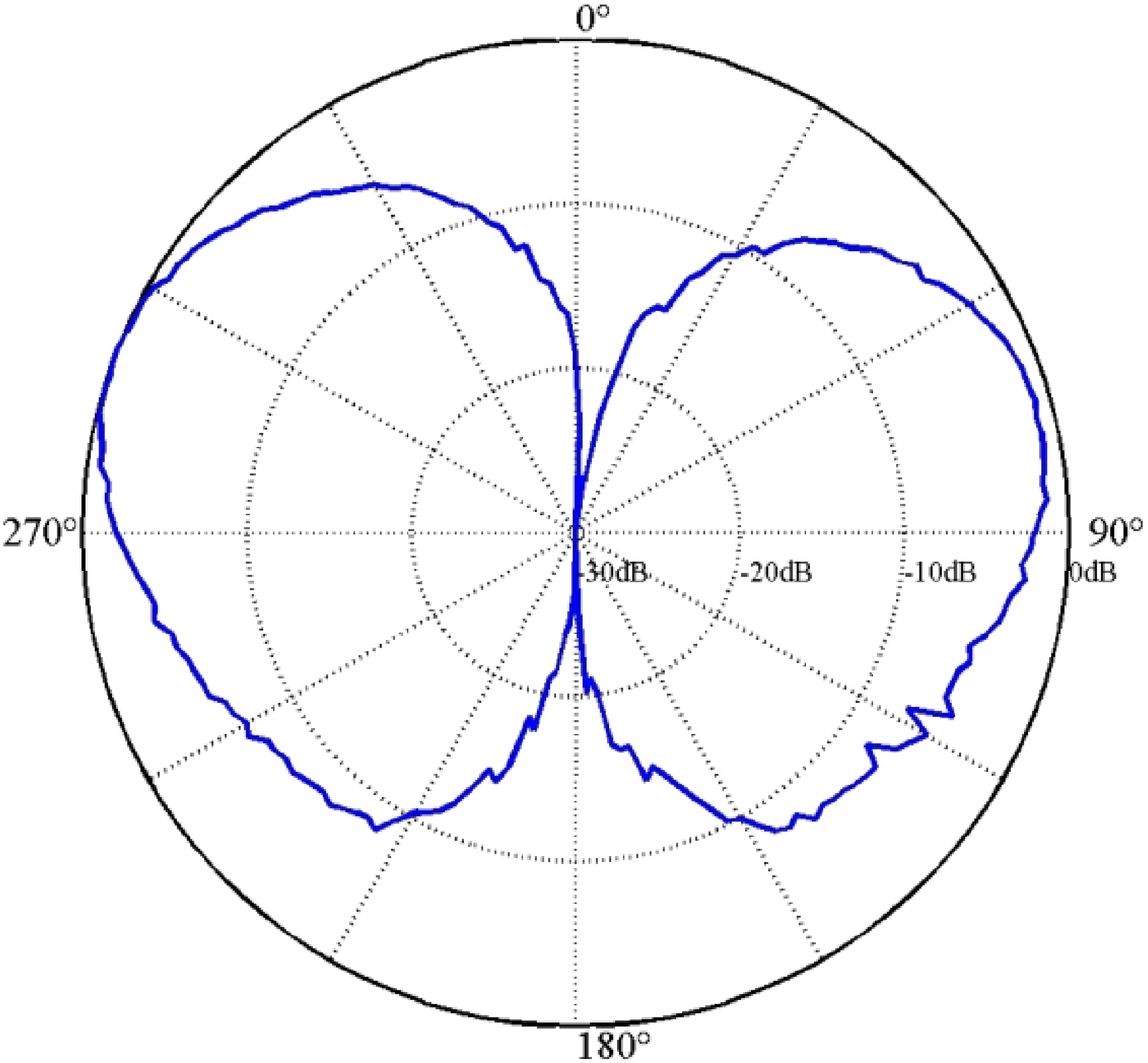}
    \captionsetup{justification=centering}
        \caption{}
        \label{fig:UWB_Antenna_Elevation_Beam_Pattern_4GHz}
    \end{subfigure}
    \caption{UWB Elevation Beam Pattern for $\SI{4}{\GHz}$ \cite{timedomain_antenna_manual}.}
    \label{fig:UWB_Antenna_Elevation_Beam_Pattern}
\end{figure}

It is evident that the accuracy in 3D real-world scenarios does not match that of simulations, partly due to the larger and more unevenly distributed measurement errors of UWB ranging sensors when measuring 3D distances. This is illustrated by the antenna beam patterns at different angles, as shown in Fig.~\ref{fig:UWB_Antenna_Azimuth_Beam_Pattern} and Fig.~\ref{fig:UWB_Antenna_Elevation_Beam_Pattern}, where the peak of the antenna gain is about $\SI{20}{\degree}$ elevation angle and about $\SI{0}{\degree}$ azimuth angle \cite{timedomain_antenna_manual}. Additionally, it is noteworthy that the RMSE often reaches or exceeds half the side length of the platform. Consequently, we arrive at a conclusion similar to that in Section~\ref{sect:3D_Simulation}: relying on UWB ranging sensors for relative attitude estimation is less effective, and using IMUs or other dedicated attitude sensors proves to be a more economical and efficient alternative. For this reason, the experiment for agent localization is omitted.

In summary, the hardware experiments provide concrete evidence of the precision, feasibility, and robustness of the proposed algorithms when faced with uncertainties and disturbances inherent in real-world 3D environments. 

\subsection{Applications}
\begin{figure}[!ht]
        \centering
        \includegraphics[width=0.4\textwidth]{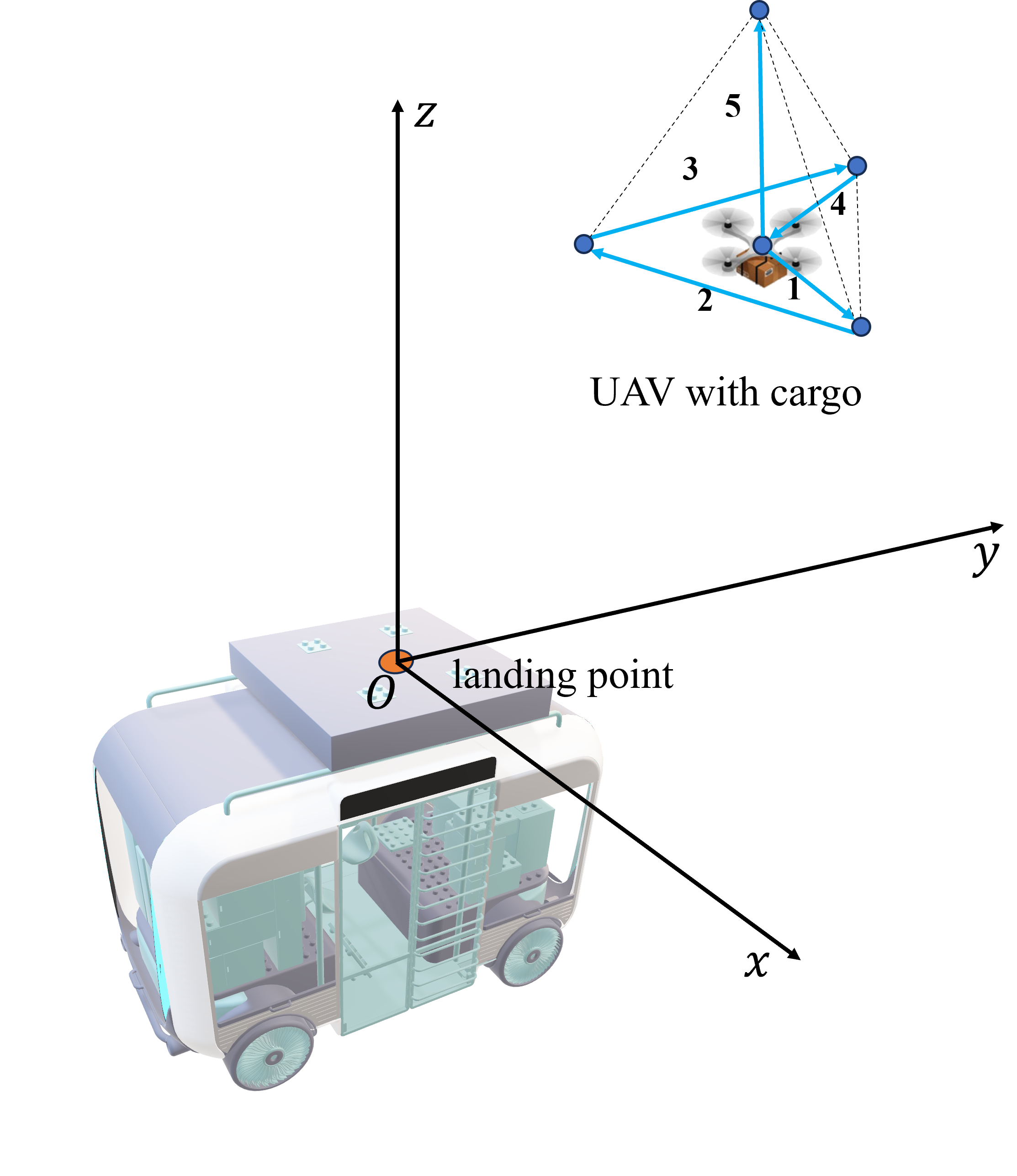}
        \caption{Near-field relative localization of UAV for cooperation with UGV truck.}
        \label{fig:UAV_UGV_Delivery_Demo}
\end{figure}

The simulation results demonstrate that the proposed 3D relative localization framework offers significant advantages for near-field 3D UxV applications. In this subsection, the potential for heterogeneous UxV cooperation and delivery is explored. Given that GNSS horizontal localization accuracy is poor in complex environments and vertical localization accuracy is even worse, there is a pressing need for a near-field localization method that ensures high accuracy in both vertical and horizontal dimensions. With only one UWB sensor configured on the UGV and UAV and with the help of IMU sensors, the UAV can move to the vertices of a tetrahedron sequentially and get its relative 3D position with respect to the UGV with a decimeter-level accuracy by utilizing our proposed framework, as demonstrated in Fig. \ref{fig:UAV_UGV_Delivery_Demo}. The localization accuracy can be improved by increasing the edge length and the face number of the regular polyhedron formed by measurement points; however, this inevitably leads to an increase in the time cost required. Therefore, the specific parameter settings should be fine-tuned based on practical considerations, including the required positioning accuracy, the actual precision of the sensors, and the operational environment.

\section{CONCLUSION}
\label{sect:Conclusion}
In this article, a 3D infrastructure-free relative localization framework is proposed with low-cost and sub-meter-level accuracy, aiming to address the issue of insufficient precision in near-field navigation under adverse environments encountered in the low-altitude economy and UAV delivery. Two 3D relative localization methods, EDMT and MLE, are proposed and compared with the state-of-the-art algorithms, TT and Fro-CVX, in terms of computational complexity and positioning accuracy. Simulations results demonstrate the low computational complexity and high sub-meter accuracy properties of the proposed methods. Field tests using UWB range-measurement sensors are conducted and the experiment results further validate the advantages of the proposed framework. 

Overall, our proposed relative localization algorithms and framework provide a solution for the improvement of the near-field navigation accuracy in UxV industry. With appropriate adjustment and extensions, our proposed framework can be further utilized to more multi-agent UxV cooperation applications in complex environments. 

\section*{ACKNOWLEDGMENT}
The authors would like to thank Axel Ringh of Chalmers University of Technology for beneficial discussions. 

\section*{REFERENCES}
\bibliographystyle{IEEEtran}
\bibliography{ReferenceCopy.bib}

\begin{thebibliography}{10}
\providecommand{\url}[1]{#1}
\csname url@samestyle\endcsname
\providecommand{\newblock}{\relax}
\providecommand{\bibinfo}[2]{#2}
\providecommand{\BIBentrySTDinterwordspacing}{\spaceskip=0pt\relax}
\providecommand{\BIBentryALTinterwordstretchfactor}{4}
\providecommand{\BIBentryALTinterwordspacing}{\spaceskip=\fontdimen2\font plus
\BIBentryALTinterwordstretchfactor\fontdimen3\font minus \fontdimen4\font\relax}
\providecommand{\BIBforeignlanguage}[2]{{%
\expandafter\ifx\csname l@#1\endcsname\relax
\typeout{** WARNING: IEEEtran.bst: No hyphenation pattern has been}%
\typeout{** loaded for the language `#1'. Using the pattern for}%
\typeout{** the default language instead.}%
\else
\language=\csname l@#1\endcsname
\fi
#2}}
\providecommand{\BIBdecl}{\relax}
\BIBdecl

\bibitem{buehrer_collaborative_2018}
R.~M. Buehrer, H.~Wymeersch, and R.~M. Vaghefi, ``Collaborative sensor network localization: Algorithms and practical issues,'' \emph{Proceedings of the IEEE}, vol. 106, no.~6, pp. 1089--1114, Jun. 2018.

\bibitem{evrendilek_complexity_2011}
C.~Evrendilek and H.~Akcan, ``On the complexity of trilateration with noisy range measurements,'' \emph{IEEE Communications Letters}, vol.~15, no.~10, pp. 1097--1099, Oct. 2011.

\bibitem{xu2025SurveyUltraWidea}
N.~Xu, M.~Guan, and C.~Wen, ``A survey on {{Ultra Wide Band}} based localization for mobile autonomous machines,'' \emph{Journal of Automation and Intelligence}, Feb. 2025.

\bibitem{madani2022HybridTruckDroneDelivery}
\BIBentryALTinterwordspacing
B.~Madani and M.~Ndiaye, ``\BIBforeignlanguage{en-US}{Hybrid {Truck}-{Drone} {Delivery} {Systems}: {A} {Systematic} {Literature} {Review}},'' \emph{\BIBforeignlanguage{en-US}{IEEE Access}}, vol.~10, pp. 92\,854--92\,878, 2022.
\BIBentrySTDinterwordspacing

\bibitem{attenni2023DroneBasedDeliverySystems}
\BIBentryALTinterwordspacing
G.~Attenni, V.~Arrigoni, N.~Bartolini, and G.~Maselli, ``\BIBforeignlanguage{en-US}{Drone-{Based} {Delivery} {Systems}: {A} {Survey} on {Route} {Planning}},'' \emph{\BIBforeignlanguage{en-US}{IEEE Access}}, vol.~11, pp. 123\,476--123\,504, 2023.
\BIBentrySTDinterwordspacing

\bibitem{moadab2022DroneRoutingProblem}
\BIBentryALTinterwordspacing
A.~Moadab, F.~Farajzadeh, and O.~Fatahi~Valilai, ``\BIBforeignlanguage{en}{Drone routing problem model for last-mile delivery using the public transportation capacity as moving charging stations},'' \emph{\BIBforeignlanguage{en}{Sci Rep}}, vol.~12, no.~1, p. 6361, Apr. 2022.
\BIBentrySTDinterwordspacing

\bibitem{WEN2022HeterogeneousMultidrone}
\BIBentryALTinterwordspacing
X.~Wen and G.~Wu, ``Heterogeneous multi-drone routing problem for parcel delivery,'' \emph{Transportation Research Part C: Emerging Technologies}, vol. 141, p. 103763, 2022.
\BIBentrySTDinterwordspacing

\bibitem{chen2025CoordinatedDroneCourierLogistics}
\BIBentryALTinterwordspacing
S.~Chen, K.~Wang, L.~Wu, and W.~Qi, ``\BIBforeignlanguage{en-US}{On {Coordinated} {Drone}-{Courier} {Logistics} for {Intra}-city {Express} {Services}},'' Jan. 2025.
\BIBentrySTDinterwordspacing

\bibitem{hahn2021SecurityPrivacyIssuesa}
D.~Hahn, A.~Munir, and V.~Behzadan, ``Security and privacy issues in intelligent transportation systems: {{Classification}} and challenges,'' \emph{IEEE Intelligent Transportation Systems Magazine}, vol.~13, no.~1, pp. 181--196, 2021.

\bibitem{bendiab2023AutonomousVehiclesSecurity}
G.~Bendiab, A.~Hameurlaine, G.~Germanos, N.~Kolokotronis, and S.~Shiaeles, ``Autonomous vehicles security: {{Challenges}} and solutions using blockchain and artificial intelligence,'' \emph{IEEE Transactions on Intelligent Transportation Systems}, vol.~24, no.~4, pp. 3614--3637, Apr. 2023.

\bibitem{gao_TIM_2025_InfrastructureFreeRelativeLocalization}
Q.~Gao, K.~Ho~Cheng, L.~Qiu, and Z.~Gong, ``Infrastructure-free relative localization: System modeling, algorithm design, performance analysis, and field tests,'' \emph{IEEE Transactions on Instrumentation and Measurement}, vol.~74, pp. 1--14, 2025.

\bibitem{zhang2022EfficientUAVLocalization}
W.~Zhang and W.~Zhang, ``An efficient {{UAV}} localization technique based on particle swarm optimization,'' \emph{IEEE Transactions on Vehicular Technology}, vol.~71, no.~9, pp. 9544--9557, Sep. 2022.

\bibitem{lu2024FastOmnidirectionalRelative}
Z.~Lu, Y.~Wu, S.~Yang, K.~Zhang, and Q.~Quan, ``Fast and omnidirectional relative position estimation with circular markers for {{UAV}} swarm,'' \emph{IEEE Transactions on Instrumentation and Measurement}, vol.~73, pp. 1--11, 2024.

\bibitem{zeng2023UAVLocalizationSystem}
Q.~Zeng, Y.~Jin, H.~Yu, and X.~You, ``A {{UAV}} localization system based on double {{UWB}} tags and {{IMU}} for landing platform,'' \emph{IEEE Sensors Journal}, vol.~23, no.~9, pp. 10\,100--10\,108, May 2023.

\bibitem{lazzari2017NumericalInvestigationUWB}
F.~Lazzari, A.~Buffi, P.~Nepa, and S.~Lazzari, ``Numerical investigation of an {{UWB}} localization technique for unmanned aerial vehicles in outdoor scenarios,'' \emph{IEEE Sensors Journal}, vol.~17, no.~9, pp. 2896--2903, May 2017.

\bibitem{mirzaeinia2020AnalyticalStudyLeader}
A.~Mirzaeinia, F.~Heppner, and M.~Hassanalian, ``An analytical study on leader and follower switching in {{V-shaped Canada Goose}} flocks for energy management purposes,'' \emph{Swarm Intell}, vol.~14, no.~2, pp. 117--141, Jun. 2020.

\bibitem{ito2022EmergenceGiantRotating}
S.~Ito and N.~Uchida, ``Emergence of a {{Giant Rotating Cluster}} of {{Fish}} in {{Three Dimensions}} by {{Local Interactions}},'' \emph{J. Phys. Soc. Jpn.}, vol.~91, no.~6, p. 064806, Jun. 2022.

\bibitem{gueron1996DynamicsHerdsIndividualsa}
S.~Gueron, S.~A. Levin, and D.~I. Rubenstein, ``The {{Dynamics}} of {{Herds}}: {{From Individuals}} to {{Aggregations}},'' \emph{Journal of Theoretical Biology}, vol. 182, no.~1, pp. 85--98, Sep. 1996.

\bibitem{de_silva_ultrasonic_2015}
\BIBentryALTinterwordspacing
O.~De~Silva, G.~K.~I. Mann, and R.~G. Gosine, ``An ultrasonic and vision-based relative positioning sensor for multirobot localization,'' \emph{IEEE Sensors Journal}, vol.~15, no.~3, pp. 1716--1726, Mar. 2015.
\BIBentrySTDinterwordspacing

\bibitem{brunacci_development_2023}
\BIBentryALTinterwordspacing
V.~Brunacci, A.~De~Angelis, G.~Costante, and P.~Carbone, ``Development and analysis of a {UWB} relative localization system,'' \emph{IEEE Transactions on Instrumentation and Measurement}, vol.~72, pp. 1--13, 2023.
\BIBentrySTDinterwordspacing

\bibitem{vasarhelyi_optimized_2018}
\BIBentryALTinterwordspacing
G.~Vásárhelyi, C.~Virágh, G.~Somorjai, T.~Nepusz, A.~E. Eiben, and T.~Vicsek, ``\BIBforeignlanguage{en}{Optimized flocking of autonomous drones in confined environments},'' \emph{\BIBforeignlanguage{en}{Science Robotics}}, vol.~3, no.~20, p. eaat3536, Jul. 2018.
\BIBentrySTDinterwordspacing

\bibitem{cao_relative_2021}
Z.~Cao, R.~Liu, C.~Yuen, A.~Athukorala, B.~K. Kiat~Ng, M.~Mathanraj, and U.-X. Tan, ``Relative localization of mobile robots with multiple ultra-wideband ranging measurements,'' in \emph{IEEE/RSJ International Conference on Intelligent Robots and Systems ({IROS})}, Sep. 2021, pp. 5857--5863.

\bibitem{fishberg_multi-agent_2022}
\BIBentryALTinterwordspacing
A.~Fishberg and J.~P. How, ``Multi-agent relative pose estimation with {UWB} and constrained communications,'' in \emph{2022 {IEEE}/{RSJ} {International} {Conference} on {Intelligent} {Robots} and {Systems} ({IROS})}, Oct. 2022, pp. 778--785.
\BIBentrySTDinterwordspacing

\bibitem{dong2023SocialSignalLearning}
S.~Dong, T.~Lin, J.~C. Nieh, and K.~Tan, ``Social signal learning of the waggle dance in honey bees,'' \emph{Science}, vol. 379, no. 6636, pp. 1015--1018, 2023.

\bibitem{shang_localization_2003}
\BIBentryALTinterwordspacing
Y.~Shang, W.~Ruml, Y.~Zhang, and M.~P.~J. Fromherz, ``Localization from mere connectivity,'' in \emph{ACM MobiHoc}, Jun. 2003, pp. 201--212.
\BIBentrySTDinterwordspacing

\bibitem{cao_sensor_2006}
\BIBentryALTinterwordspacing
M.~Cao, B.~D.~O. Anderson, and A.~S. Morse, ``\BIBforeignlanguage{en}{Sensor network localization with imprecise distances},'' \emph{\BIBforeignlanguage{en}{Systems \& Control Letters}}, vol.~55, no.~11, pp. 887--893, Nov. 2006.
\BIBentrySTDinterwordspacing

\bibitem{cao_formation_2011}
\BIBentryALTinterwordspacing
M.~Cao, C.~Yu, and B.~D.~O. Anderson, ``\BIBforeignlanguage{en}{Formation control using range-only measurements},'' \emph{\BIBforeignlanguage{en}{Automatica}}, vol.~47, no.~4, pp. 776--781, Apr. 2011.
\BIBentrySTDinterwordspacing

\bibitem{cheok_uwb_2010}
\BIBentryALTinterwordspacing
K.~C. Cheok, M.~Radovnikovich, P.~Vempaty, G.~R. Hudas, J.~L. Overholt, and P.~Fleck, ``{UWB} tracking of mobile robots,'' in \emph{21st {Annual} {IEEE} {International} {Symposium} on {Personal}, {Indoor} and {Mobile} {Radio} {Communications}}, Sep. 2010, pp. 2615--2620.
\BIBentrySTDinterwordspacing

\bibitem{thomas_revisiting_2005}
F.~Thomas and L.~Ros, ``Revisiting trilateration for robot localization,'' \emph{IEEE Transactions on Robotics}, vol.~21, no.~1, pp. 93--101, Feb. 2005.

\bibitem{li_novel_2017}
\BIBentryALTinterwordspacing
J.~Li, X.~Yue, J.~Chen, and F.~Deng, ``\BIBforeignlanguage{en}{A novel robust trilateration method applied to ultra-wide bandwidth location systems},'' \emph{\BIBforeignlanguage{en}{Sensors}}, vol.~17, no.~4, p. 795, Apr. 2017.
\BIBentrySTDinterwordspacing

\bibitem{so_theory_2007}
\BIBentryALTinterwordspacing
A.~M.-C. So and Y.~Ye, ``\BIBforeignlanguage{en}{Theory of semidefinite programming for {Sensor} {Network} {Localization}},'' \emph{\BIBforeignlanguage{en}{Mathematical Programming}}, vol. 109, no.~2, pp. 367--384, Mar. 2007.
\BIBentrySTDinterwordspacing

\bibitem{dattorro_convex_2019}
\BIBentryALTinterwordspacing
J.~Dattorro, \emph{\BIBforeignlanguage{en}{Convex {Optimization} \& {Euclidean} {Distance} {Geometry} 2$\varepsilon$}}.\hskip 1em plus 0.5em minus 0.4em\relax Palo Alto, CA: MEBOO Publishing, Oct. 2019.
\BIBentrySTDinterwordspacing

\bibitem{chu_least_2008}
D.~I. Chu, H.~C. Brown, and M.~T. Chu, ``On least squares {Euclidean} distance matrix approximation and completion,'' Department of Mathematics, North Carolina State University, techreport, 2003.

\bibitem{gao_euclidean_2019}
\BIBentryALTinterwordspacing
J.~Fliege, H.-D. Qi, and N.~Xiu, ``\BIBforeignlanguage{en}{Euclidean distance matrix optimization for sensor network localization*},'' in \emph{\BIBforeignlanguage{en}{Cooperative {Localization} and {Navigation}}}, 1st~ed., C.~Gao, G.~Zhao, and H.~Fourati, Eds.\hskip 1em plus 0.5em minus 0.4em\relax CRC Press, Aug. 2019, pp. 99--126.
\BIBentrySTDinterwordspacing

\bibitem{menger_new_1931}
\BIBentryALTinterwordspacing
K.~Menger, ``New foundation of {Euclidean} geometry,'' \emph{American Journal of Mathematics}, vol.~53, no.~4, pp. 721--745, Oct. 1931.
\BIBentrySTDinterwordspacing

\bibitem{schoenberg_remarks_1935}
\BIBentryALTinterwordspacing
I.~J. Schoenberg, ``Remarks to {Maurice} {Frechet}'s article ``{Sur} la definition axiomatique {D}'une classe d'espace distances vectoriellement applicable sur l'espace de {Hilbert},'' \emph{Annals of Mathematics}, vol.~36, no.~3, pp. 724--732, Jul. 1935.
\BIBentrySTDinterwordspacing

\bibitem{mathar_best_1985}
\BIBentryALTinterwordspacing
R.~Mathar, ``The best {Euclidian} fit to a given distance matrix in prescribed dimensions,'' \emph{Linear Algebra and its Applications}, vol.~67, pp. 1--6, Jun. 1985.
\BIBentrySTDinterwordspacing

\bibitem{strang_linear_2019}
G.~Strang, \emph{\BIBforeignlanguage{English}{Linear {Algebra} and {Learning} from {Data}}}.\hskip 1em plus 0.5em minus 0.4em\relax Wellesley, MA: Wellesley-Cambridge Press, Jan. 2019.

\bibitem{eggert_estimating_1997}
\BIBentryALTinterwordspacing
D.~Eggert, A.~Lorusso, and R.~Fisher, ``Estimating 3-{D} rigid body transformations: a comparison of four major algorithms,'' \emph{Machine Vision and Applications}, vol.~9, no.~5, pp. 272--290, Mar. 1997.
\BIBentrySTDinterwordspacing

\bibitem{kay_fundamentals_1993}
S.~Kay, \emph{\BIBforeignlanguage{English}{Fundamentals of {Statistical} {Signal} {Processing}, {Volume} {I}: {Estimation} {Theory}}}, 1st~ed.\hskip 1em plus 0.5em minus 0.4em\relax Englewood Cliffs, NJ: Pearson, Apr. 1993.

\bibitem{timedomain_antenna_manual}
\emph{{Broadspec} {UWB} {Antenna} {TDSR} {UWB} {Radios}}, TDSR LLC, 2020.

\end{thebibliography}

\end{document}